\documentclass[11pt]{article}
\usepackage{a4wide}
\usepackage{amsmath,amssymb}
\usepackage{graphicx}
\usepackage{color}
\usepackage{braket}
\usepackage{comment}
\usepackage[bookmarks=true,bookmarksnumbered=true,setpagesize=false]{hyperref}

\DeclareMathOperator{\sgn}{sgn}
\DeclareMathOperator{\erf}{erf}
\DeclareMathOperator{\T}{T}
\DeclareMathOperator{\diag}{diag}

\renewcommand{\t}{\text{t}}

\usepackage{enumerate}
\newcommand{\al}[1]{\begin{align}#1\end{align}}
\newcommand{\als}[1]{\begin{align*}#1\end{align*}}
\newcommand{\ov}{\over}
\newcommand{\nn}{\nonumber\\}
\newcommand{\tx}{\text}

\newcommand{\paren}[1]{\left(#1\right)}
\newcommand{\sqbr}[1]{\left[#1\right]}
\newcommand{\ab}[1]{\left|#1\right|}

\newcommand{\fn}[1]{\!\left(#1\right)}

\newcommand{\bs}{\boldsymbol}

\newcommand{\red}[1]{#1}

\newcommand{\blue}[1]{#1}

\newcommand{\df}{\text{d}}

\newcommand{\mc}{\mathcal}
\newcommand{\mf}{\mathfrak}
\newcommand{\bmf}[1]{\boldsymbol{\mathfrak#1}}
\newcommand{\bmat}[1]{\begin{bmatrix}#1\end{bmatrix}}

\newcommand{\p}{\partial}

\newcommand{\Or}[1]{\mathcal O\!\left(#1\right)}
\newcommand{\h}{\hat}
\newcommand{\ol}{\overline}
\newcommand{\pr}{\prime}

\newcommand{\commutator}[2]{\left[#1,#2\right]}

\newcommand{\wt}{\widetilde}

\newcommand{\ora}{\overrightarrow}

\newcommand{\Hp}{\tx{(H)}}
\newcommand{\Sp}{\tx{(S)}}
\newcommand{\Ip}{\tx{(I)}}

\newcommand{\Hpr}{\!{}^\tx{(H)}\,}
\newcommand{\Spr}{\!{}^\tx{(S)}\,}
\newcommand{\Ipr}{\!{}^\tx{(I)}\,}
\newcommand{\SBr}{\!{}^\tx{(SB)}\,}
\newcommand{\IBr}{\!{}^\tx{(IB)}\,}
\newcommand{\HBr}{\!{}^\tx{(HB)}\,}

\newcommand{\SBrP}{\!{}^\tx{(SB)}_{\Psi}\,}
\newcommand{\IBrP}{\!{}^\tx{(IB)}_{\Psi}\,}

\newcommand{\SBrPd}{\!{}^\tx{(SB)}_{\Psi'}\,}
\newcommand{\IBrPd}{\!{}^\tx{(IB)}_{\Psi'}\,}

\newcommand{\SBlP}{{}^\tx{\ (SB)}_{\ \ \ \,\Psi}\!}
\newcommand{\IBlP}{{}^\tx{\ (IB)}_{\ \ \ \Psi}\!}

\newcommand{\HprPh}{\!{}^\tx{(H)}_{\Phi}\,}

\newcommand{\IprPh}{\!{}^\tx{(I)}_{\Phi}\,}
\newcommand{\SBrPh}{\!{}^\tx{(SB)}_{\Phi}\,}
\newcommand{\IBrPh}{\!{}^\tx{(IB)}_{\Phi}\,}

\newcommand{\SBlPh}{\,{}^\tx{(SB)}_{\ \ \ \,\Phi}\!\!}

\newcommand{\Hprpp}{\!{}^\tx{(H)}_{\phi\phi}\,}

\newcommand{\Iprpp}{\!{}^\tx{(I)}_{\phi\phi}\,}
\newcommand{\SBrpp}{\!{}^\tx{(SB)}_{\phi\phi}\,}

\newcommand{\SBrp}{\!{}^\tx{(SB)}_{\phi}\,}
\newcommand{\IBrp}{\!{}^\tx{(IB)}_{\phi}\,}

\newcommand{\Hplpp}{\,{}^\tx{(H)}_{\phi\phi}\!\!}

\newcommand{\SBlpp}{{}^\tx{\ (SB)}_{\ \ \,\phi\phi}\!}

\newcommand{\SBlp}{\,{}^\tx{(SB)}_{\ \ \ \,\phi}\!\!}

\begin{document}
\title{Particle decay in Gaussian wave-packet formalism revisited}
\author{
Kenzo~Ishikawa\footnote{
E-mail: \tt ishikawa@particle.sci.hokudai.ac.jp}
\
 and
Kin-ya~Oda\footnote{
E-mail: \tt odakin@phys.sci.osaka-u.ac.jp
} 
}
\maketitle
\begin{center}
$^*$ \it Department of Physics, 
Hokkaido University, Hokkaido 060-0810, Japan \\
$^*$ \it Research and Education Center for Natural Sciences, Keio University\\
Kanagawa 223-8521, Japan\\
$^\dagger$ \it Department of Physics, 
Osaka University, Osaka 560-0043, Japan 
\end{center}


\begin{abstract}\noindent
We derive the Fermi's golden rule in the Gaussian wave-packet formalism of quantum field theory, proposed by Ishikawa, Shimomura, and Tobita, for the particle decay within a finite time interval.
We present a systematic procedure to separate the bulk contribution from those of time boundaries, while manifestly maintaining the unitarity of the $S$-matrix unlike the proposal by Stueckelberg in 1951.
We also revisit the suggested deviation from the golden rule and clarify that it indeed corresponds to the boundary contributions, though their physical significance is yet to be confirmed.
\end{abstract}

\vfill

\begin{flushright}
OU-HET/985\\
EPHOU-18-013
\end{flushright}
\newpage

\tableofcontents

\newpage

\section{Introduction}
Strictly speaking, the $S$-matrix in quantum field theory is defined only by using wave packets; see any textbook, e.g., Ref.~\cite{Peskin:1995ev,Weinberg:1995mt}. The derivation of a physical quantity, such as a decay rate, in terms of plane waves is \emph{``actually more a mnemonic than a derivation''}~\cite{Weinberg:1995mt}.

Ishikawa and Shimomura have proposed a formulation of a free Gaussian wave packet in relativistic quantum field theory~\cite{Ishikawa:2005zc}; see also Refs.~\cite{doi:10.1143/PTP.17.419,doi:10.1143/PTP.18.101b,doi:10.1143/PTP.20.857,Cho:1993xe} for earlier related works.
Ishikawa and Tobita have developed a systematic method to approximate the $S$-matrix in various limits in the Gaussian wave-packet formalism~\cite{Ishikawa:2014eoa,Ishikawa:2013kba,Ishikawa:2016lnn}; further development has been made by themselves and Tajima to include the photon state~\cite{Ishikawa:2014uma}. The authors have claimed that there can be a deviation from the Fermi's golden rule if we consider an $S$-matrix with finite time interval~\cite{Ishikawa:2014eoa,Ishikawa:2013kba,Ishikawa:2014uma,Ishikawa:2016lnn}.

Stueckelberg correctly pointed out in 1951 that the plane-wave $S$-matrix with finite time interval exhibits an extra ultraviolet (UV) divergence coming from the interaction point at the boundary in time~\cite{Stueckelberg:1951zz}:
In order to remove it within the plane-wave formalism, a phenomenological factor has been introduced so that the uncertainty of the initial and final times of the process can be taken into account. This has lead to the violation of unitarity, and the necessary modification of the $S$-matrix to cure the pathology has become complicated and rather intractable.

In this paper, we revisit the Gaussian wave-packet formalism to \emph{derive} the Fermi's golden rule.
We separate the bulk effect from the boundary ones, while manifestly maintaining the unitarity.
We further show that the might-be deviation from the Fermi's golden rule, claimed in Refs.~\cite{Ishikawa:2014eoa,Ishikawa:2013kba,Ishikawa:2014uma,Ishikawa:2016lnn}, indeed corresponds to the decay at the boundary in time.

For clarity, in Secs.~\ref{Gaussian section}--\ref{decay probability section}, we will first spell out our results using an example of the tree-level decay process of a heavy scalar $\Phi$ into a pair of light scalars $\phi\phi$ due to the super-renormalizable interaction~$\Phi\phi\phi$.
In order to show how to generalize our results to include the momentum-dependent factors in the interaction and in the wave functions, in Sec.~\ref{diphoton section}, we will then turn to the tree-level decay process of a pseudo-scalar $\varphi$ into a pair of photons due to the non-renormalizable interaction $\varphi F_{\mu\nu}\tilde F^{\mu\nu}$. More generalization will be presented in Appendix~\ref{GWP section}.

The paper is organized as follows:
In Sec.~\ref{Gaussian section}, we review the Gaussian wave-packet formalism for the scalar field.
In Sec.~\ref{S-matrix section}, we reformulate the Gaussian $S$-matrix and present a systematic procedure to separate the bulk contribution from the boundary ones.
In Sec.~\ref{decay probability section}, we obtain the decay probability and derive the Fermi's golden rule. We briefly discuss the boundary effect too.
In Sec.~\ref{diphoton section}, we generalize our result to the decay into the diphoton final state.
In Sec.~\ref{summary}, we summarize our results.
In Appendix~\ref{GWP section}, we review the Gaussian wave-packet formalism for the scalar, spinor, and vector.
In Appendix~\ref{saddle point section}, we show the saddle-point approximation of the Gaussian wave packet in the large-width (plane-wave) expansion.
In Appendix~\ref{limits section}, we show the expressions for the plane-wave and particle limits of the decaying particle and for the decay at rest.
In Appendix~\ref{Boundary limit}, we present possible expressions 
for the boundary limit. 

\section{Gaussian formalism}\label{Gaussian section}
We review the Gaussian formalism. As said above, we consider the decay of a heavy real scalar $\Phi$ into a pair of light real scalars $\phi\phi$ by the following interaction:
\al{
\mc L_\tx{int}
	&=	-{\kappa\ov2}\Phi\phi^2,
		\label{superrenormalizable interaction}
}
where $\kappa$ is a coupling constant of mass dimension unity.
The interaction Hamiltonian density is $\mc H_\tx{int}=-\mc L_\tx{int}$.
We write the initial and final momenta $\bs p_0$ and $\bs p_1,\bs p_2$, respectively.
In this section, we will let $\Psi$ stand for either $\Phi$ or $\phi$.
We write their masses $m_\Phi$ and $m_\phi$ and consider the case $m_\Phi>2m_\phi$.

\subsection{Plane-wave $S$-matrix}
First we briefly review the plane-wave computation of the $S$-matrix.
We can expand the free field operator $\h\Psi^\Ip\fn{x}$ at $x=\paren{x^0,\bs x}=\paren{t,\bs x}$ in the interaction picture in terms of the annihilation and creation operators of  planes waves:
\al{
\h\Psi^\Ip\fn{x}
	&=	\left.
			\int
			{\df^3\bs p\ov\sqrt{2p^0}\paren{2\pi}^{3/2}}
			\sqbr{
				\h a_\Psi\fn{\bs p}e^{ip\cdot x}
				+\h a_\Psi^\dagger\fn{\bs p}e^{-ip\cdot x}
				}
		\,
		\right|_{p^0=E_\Psi\fn{\bs p}},
		\label{plane wave expansion}
}
where we work in the $(-,+,+,+)$ metric convention and write the kinetic energy
\al{
E_\Psi\fn{\bs p}
	:=\sqrt{m_\Psi^2+\bs p^2}.
		\label{on-shell energy}
}
Throughout this paper, we use both $x^0$ and $t$ interchangeably (as well as $X^0$ and $T$ that appear below).

We define the following free one- and two-particle states:\footnote{
The two-particle state is normalized to
\als{
\SBlpp\Braket{\bs p_1,\bs p_2|\bs p_3,\bs p_4}\SBrpp
	&=	{1\ov2}\sqbr{\delta^3\fn{\bs p_1-\bs p_3}\delta^3\fn{\bs p_2-\bs p_4}+\delta^3\fn{\bs p_1-\bs p_4}\delta^3\fn{\bs p_2-\bs p_3}},
}
such that
\als{
\int\df^3\bs p_1\int\df^3\bs p_2\Ket{\bs p_1,\bs p_2}\Bra{\bs p_1,\bs p_2}
	&=	\h 1,
}
where $\h 1$ is the identity operator in the two-particle subspace.
}
\al{
\Ket{\bs p_0}\SBrPh
	&=	\h a_\Phi^\dagger\fn{\bs p_0}\Ket{0},\nn
\Ket{\bs p}\SBrp
	&=	\h a_\phi^\dagger\fn{\bs p}\Ket{0},\nn
\Ket{\bs p_1,\bs p_2}\SBrpp
	&=	{1\ov\sqrt{2}}\h a_\phi^\dagger\fn{\bs p_1}\h a_\phi^\dagger\fn{\bs p_2}\Ket{0},
		\label{free states}
}
where $\tx{(SB)}$ refers to the time-independent basis state in the Schr\"odinger picture (see Appendix~\ref{pictures}), which are the eigenstates of the free Hamiotonian:
\al{
\h H_\tx{free}\Ket{\bs p_0}\SBrPh
	&=	E_\Phi\fn{\bs p_0}\Ket{\bs p_0}\SBrPh, \nn
\h H_\tx{free}\Ket{\bs p}\SBrp
	&=	E_\phi\fn{\bs p}\Ket{\bs p}\SBrp, \nn
\h H_\tx{free}\Ket{\bs p_1,\bs p_2}\SBrpp
	&=	\paren{E_\phi\fn{\bs p_1}+E_\phi\fn{\bs p_2}}\Ket{\bs p_1,\bs p_2}\SBrpp.
}
In terms of these states, the free field operator~\eqref{plane wave expansion} can also be written as
\al{
\h\Psi^\Ip\fn{x}
	&=	\int
			{\df^3\bs p\ov\sqrt{2E_\Psi\fn{\bs p}}}
			\sqbr{
				\h a_\Psi\fn{\bs p}
				\paren{\IBlP\Braket{x|\bs p}\SBrP}
				+\h a_\Psi^\dagger\fn{\bs p}
				\paren{\IBlP\Braket{x|\bs p}\SBrP}^*
				},
}
where $\Ket{x}\IBrP=e^{i\h H_\tx{free}t}\Ket{\bs x}\SBrP$ is the position basis state in the interaction picture; see Appendix~\ref{plane wave expansion section}.

Usually, the time-independent in and out states in the Heisenberg picture are defined as the eigenstates of the total Hamiltonian that become close to the free states~\eqref{free states} at sufficiently remote past and future in the following sense:\footnote{
This can be formally rewritten as the interaction-picture state becoming close to the time-independent Schr\"odinger basis state as
\als{
\Ket{\tx{in};\bs p_0,t}\IprPh
	&\to	\Ket{\bs p_0}\SBrPh,&
\tx{for }t
	&\to	T_\tx{in}\tx{ ($\to-\infty$),}\\
\Ket{\tx{out};\bs p_1,\bs p_2;t}\Iprpp
	&\to	\Ket{\bs p_1,\bs p_2}\SBrpp,&
\tx{for }t 
	&\to	T_\tx{out}\tx{ ($\to\infty$).}
}
}
\al{
e^{-i\h Ht}\Ket{\tx{in};\bs p_0}\HprPh
	&\to	e^{-i\h H_\tx{free}t}\Ket{\bs p_0}\SBrPh&
\tx{for }t
	&\to	T_\tx{in} \tx{ ($\to-\infty$),} \label{in identification}\\
e^{-i\h Ht}\Ket{\tx{out};\bs p_1,\bs p_2}\Hprpp
	&\to	e^{-i\h H_\tx{free}t}\Ket{\bs p_1,\bs p_2}\SBrpp&
\tx{for }t
	&\to	T_\tx{out} \tx{ ($\to\infty$),}\label{out identification}
}
where $\h H=\h H_\tx{free}+\h H_\tx{int}$ is the total Hamiltonian.
To be more precise, \red{Eqs.~\eqref{in identification} and \eqref{out identification} are meaningless in themselves and should rather be understood as follows (see any textbook, e.g., Refs.~\cite{Peskin:1995ev,Weinberg:1995mt}): The in and out states}
are \red{really} defined by \emph{wave packets} such that\red{,} for arbitrary smooth and sufficiently fast-decaying functions $g_\tx{in}\fn{\bs p_0}$ and $g_\tx{out}\fn{\bs p_1,\bs p_2}$, they satisfy
\al{
\int\df^3\bs p_0\,g_\tx{in}\fn{\bs p_0}\,
e^{-i\h Ht}\Ket{\tx{in};\bs p_0}\HprPh
	&\approx	\int\df^3\bs p_0\,g_\tx{in}\fn{\bs p_0}\,
			e^{-i\h H_\tx{free}t}\Ket{\bs p_0}\SBrPh,
				\label{in wave packet}\\
\int\df^3\bs p_1\,\df^3\bs p_2\,g_\tx{out}\fn{\bs p_1,\bs p_2}\,
e^{-i\h Ht}\Ket{\tx{out};\bs p_1,\bs p_2}\Hprpp
	&\approx	\int\df^3\bs p_1\,\df^3\bs p_2\,g_\tx{out}\fn{\bs p_1,\bs p_2}\,
			e^{-i\h H_\tx{free}t}\Ket{\bs p_1,\bs p_2}\SBrpp,
				\label{out wave packet}
}
as $t\to T_\tx{in}$ ($\to-\infty$) and $t\to T_\tx{out}$ ($\to\infty$), respectively.\footnote{
Strictly speaking, this cannot apply for a decay process:
No matter how remote past we move on to, $t\to T_\tx{in}$ ($\to-\infty$), we might still find a wave-packet configuration of the final-state particles in which we cannot neglect the interaction at the initial time. To handle this issue, one needs to treat the production process of the parent particle using wave packets too. This will be presented in a separate publication.
\label{cannot dump interaction}
}

The $S$-matrix is defined by
\al{
S	&=	\Hplpp\Braket{\tx{out};\bs p_1,\bs p_2|\tx{in};\bs p_0}\HprPh.
	\label{plane wave S-matrix}
}
For $T_\tx{in}$ and $T_\tx{out}$ sufficiently remote past and future, respectively, one obtains
\al{
S	&\approx
		\SBlpp\Bra{\bs p_1,\bs p_2}\h U\fn{T_\tx{out},T_\tx{in}}\Ket{\bs p_0}\SBrPh,
		\label{plane-wave S-matrix}
}
where
\al{
\h U\fn{T_\tx{out},T_\tx{in}}
	&:=	e^{i\h H_\tx{free}T_\tx{out}}e^{-i\h H\paren{T_\tx{out}-T_\tx{in}}}e^{-i\h H_\tx{free}T_\tx{in}}\nn
	&=	\T\exp\fn{-i\int_{T_\tx{in}}^{T_\tx{out}}\df t\,\h H_\tx{int}^\Ip\fn{t}},
		\label{Dyson series}
}
in which $\T$ denotes the time-ordering and $\h H_\tx{int}^\Ip\fn{t}=e^{i\h H_\tx{free}t}\h H_\tx{int}e^{-i\h H_\tx{free}t}$ is the interaction Hamiltonian in the interaction picture.

As is well known, the expression~\eqref{plane-wave S-matrix} is badly divergent \red{when squared, being proportional to the momentum-space delta function $\delta^4\fn{0}$. Also,} one needs to insert an infinitesimal imaginary part for the interaction Hamiltonian by hand \red{in order to make the perturbation~\eqref{Dyson series} convergent}. This is because the overlap between plane waves can never be suppressed no matter how remote past and future one moves on, which is the reason why one needs wave packets\red{~\eqref{in wave packet} and \eqref{out wave packet}} for complete treatment of the $S$-matrix. \red{The cluster decomposition never occurs for the infinitely spread plane waves, while it does for properly defined wave packets.}

\subsection{Gaussian basis}
Now we switch from the plane-wave basis to the Gaussian basis.
Detailed notations for this subsection can be found in Appendix~\ref{GWP section}.

Instead of the plane-wave expansion~\eqref{plane wave expansion}, one may also expand the free field in terms of the annihilation and creation operators of the free Gaussian wave: 
\al{
\h\Psi^\Ip\fn{x}
	&=	\int{\df^3\bs X\,\df^3\bs P\ov\paren{2\pi}^3}
		\sqbr{
			f_{\Psi,\sigma;X,\bs P}\fn{x}\h A_{\Psi,\sigma}\fn{X,\bs P}
			+f_{\Psi,\sigma;X,\bs P}^*\fn{x}\h A_{\Psi,\sigma}^\dagger\fn{X,\bs P}
			},
}
where $\sqrt{\sigma}$ is the width of the wave packet;
$\bs X$ is the location of center at time $T$ (and we write collectively $X=\paren{X^0,\bs X}=\paren{T,\bs X}$ as said above);
 and 
 $\bs P$ is its central momentum. 
We also use the shorthand notation
\al{
\Pi	&:=	\paren{X,\bs P},&
\bs\Pi	&:=	\paren{\bs X,\bs P},&
\df^6\bs\Pi
	&:=	{\df^3\bs X\,\df^3\bs P\ov\paren{2\pi}^3},
		\label{shorthand notation}
}
so that
\al{
\h\Psi^\Ip\fn{x}
	&=	\int\df^6\bs\Pi
			\sqbr{
				f_{\Psi,\sigma;\Pi}\fn{x}\h A_{\Psi,\sigma}\fn{\Pi}
				+f_{\Psi,\sigma;\Pi}^*\fn{x}\h A_{\Psi,\sigma}^\dagger\fn{\Pi}
				}.
}
The explicit form of the coefficient function $f_{\Psi,\sigma;\Pi}$ ($=f_{\Psi,\sigma;X,\bs P}$) is obtained as\footnote{
Note that the two ``interaction basis'' states are the ones at different times:
\als{
\Ket{x}\IBrP
	&=	e^{i\h H_\tx{free}t}\Ket{\bs x}\SBrP,	&
\Ket{\Pi}\IBrP
	&=	e^{i\h H_\tx{free}T}\Ket{\bs\Pi}\SBrP,
}
where $t=x^0$ and $T=X^0$ in $\Pi=\paren{X,\bs P}$ as always.
}
\al{
f_{\Psi,\sigma;\Pi}\fn{x}
	&=	\int{\df^3\bs p\ov\sqrt{2E_\Psi\fn{\bs p}}}\IBlP\Braket{x|\bs p}\SBrP\SBlP\Braket{\bs p|\Pi}\IBrP\nn
	&=	\paren{\sigma\ov\pi}^{3/4}
		\left.
		\int{\df^3\bs p\ov\sqrt{2p^0}\paren{2\pi}^{3/2}}
		e^{ip\cdot\paren{x-X}-{\sigma\ov2}\paren{\bs p-\bs P}^2}
		\right|_{p^0=E_\Psi\fn{\bs p}}.
}
Throughout the main text, we abbreviate e.g.\ $\Ket{\sigma;\Pi}$ to $\Ket{\Pi}$, in which it is understood that $\sigma$ can be different from each other among the in- and out-state particles.

In the large-$\sigma$ expansion, the leading saddle point approximation gives
\al{
f_{\Psi,\sigma;\Pi}\fn{x}
	&\to
		\left.
		\paren{\sigma\ov\pi}^{3/4}
		\paren{2\pi\ov\sigma}^{3/2}
		{1\ov\sqrt{2P^0}\paren{2\pi}^{3/2}}e^{iP\cdot\paren{x-X}-{\paren{\bs x-\bs \Xi\fn{t}}^2\ov2\sigma}}
		\right|_{P^0=E_\Psi\fn{\bs P}},
		\label{plane wave limit}
}
where
\al{
\bs \Xi\fn{t}
	:=	\bs X+\bs V_\Psi\fn{\bs P}\paren{t-T}
		\label{delta x defined}
}
is the location of the center of the wave packet at time $t$, in which $\bs V_\Psi\fn{\bs P}:=\bs P/E_\Psi\fn{\bs P}$;
see Appendix~\ref{saddle point section}.\footnote{
$\bs \Xi\fn{t}$ has implicit dependence on  $m_\Psi$, $\bs P$, and $X$ ($=\paren{T,\bs X}$).
}
Within this leading order approximation, the width of the wave pack remains constant in time.

\subsection{Free Gaussian wave-packet states}
Now we can explicitly prepare the free wave-packet states, employed in the right-hand sides of Eqs.~\eqref{in wave packet} and \eqref{out wave packet},
\al{
&\int\df^3\bs p_0\,g_\tx{in}\fn{\bs p_0}\Ket{\bs p_0}\SBrPh,	&
&\int\df^3\bs p_1\int\df^3\bs p_2\,g_\tx{out}\fn{\bs p_1,\bs p_2}\Ket{\bs p_1,\bs p_2}\SBrpp,
}
respectively, as follows:\footnote{
Explicitly, $g_\tx{in}$ ($g_\tx{out}$) is a (multiple of independent) free Gaussian wave function(s):
\als{
g_\tx{in}\fn{\bs p}
	&=	\SBlPh\Braket{\bs p|\Pi}\IBrPh
	=	\left.
		\paren{\sigma\ov\pi}^{3/4}e^{-ip\cdot X}e^{-{\sigma\ov2}\paren{\bs p-\bs P}^2}
		\right|_{p^0=E_\Phi\fn{\bs p}},\\
g_\tx{out}\fn{\bs p_1,\bs p_2}
	&=	\prod_{a=1,2}\SBlp\Braket{\bs p_a|\Pi_a}\IBrp
	=	\prod_{a=1,2}\left.\paren{\sigma_a\ov\pi}^{3/4}e^{-ip_a\cdot X_a}e^{-{\sigma_a\ov2}\paren{\bs p_a-\bs X_a}^2}\right|_{p_a^0=E_\phi\fn{\bs p_a}},
}
where each ``interaction basis'' state is the one at different time: $\Ket{\Pi_A}\IBrP=e^{i\h H_\tx{free}T_A}\Ket{\bs\Pi_A}\SBrP$.
Note also that we have written the states in Eq.~\eqref{Gaussian free states} as the time-independent Schr\"odinger basis states, rather than the interaction basis ones, in the sense that they are independent of the time coordinate $t$ that will appear later in $\h H^\Ip\fn{t}$, the interaction Hamiltonian in the interaction picture. (Otherwise the two-particle state would have two reference times $T_1$ and $T_2$ meaninglessly.)
}
\al{
\Ket{\Pi}\SBrPh
	&=	\h A_\Phi^\dagger\fn{\Pi}\Ket{0},	&
\Ket{\Pi_1,\Pi_2}\SBrpp
	&=	{1\ov\sqrt{2}}\h A_\phi^\dagger\fn{\Pi_1}\h A_\phi^\dagger\fn{\Pi_2}\Ket{0}.
	\label{Gaussian free states}
}
As said above, $\Ket{\sigma_1,\Pi_1;\sigma_2,\Pi_2}$ is abbreviated to $\Ket{\Pi_1,\Pi_2}$ throughout the main text.

\subsection{Gaussian $S$-matrix}
Suppose that the interaction~\eqref{superrenormalizable interaction} is negligible at some initial and final times $T_\tx{in}$ and $T_\tx{out}$. Then we may define the corresponding in and out states, following Eqs.~\eqref{in wave packet} and \eqref{out wave packet}, by\footnote{
If we may take $T_0=T_\tx{in}$ and $T_1=T_2=T_\tx{out}$, we would obtain
\als{
e^{-i\h Ht}\Ket{\tx{in};\Pi_0}\HprPh
	&\approx	\Ket{\bs\Pi_0}\SBrPh&
	&(t\to T_\tx{in}),\\
e^{-i\h Ht}\Ket{\tx{out};\Pi_1,\Pi_2}\Hprpp
	&\approx	\Ket{\bs\Pi_1,\bs\Pi_2}\SBrpp&
	&(t\to T_\tx{out}),
}
respectively.
}
\al{
e^{-i\h Ht}\Ket{\tx{in};\Pi_0}\HprPh
	&\approx
e^{-i\h H_\tx{free}t}\Ket{\Pi_0}\SBrPh&
	&	(t\to T_\tx{in}),	\nn
e^{-i\h Ht}\Ket{\tx{out};\Pi_1,\Pi_2}\Hprpp
	&\approx	
e^{-i\h H_\tx{free}t}\Ket{\Pi_1,\Pi_2}\SBrpp&
	&	(t\to T_\tx{out}).
		\label{in and out states for scalar decay}
}
Now the Gaussian $S$-matrix is the inner product between these physical states:
\al{
S
	&=	\Hplpp\Braket{\tx{out};\Pi_1,\Pi_2|\tx{in};\Pi_0}\HprPh.
			\label{scalar decay S-matrix}
}
Note that these in and out states become close, in the sense of Eq.~\eqref{in and out states for scalar decay}, to the free states~\eqref{Gaussian free states}, which are square-integrable and of finite norm.\footnote{
Note however the issue in footnote~\ref{cannot dump interaction}.
}
This is in contrast to the plane-wave $S$-matrix~\eqref{plane wave S-matrix}, which is the inner product between the states that become close to the plane waves~\eqref{free states}, which are not square-integrable, not elements of the Hilbert space, and hence not the physical states.\footnote{
One can extend the notion of Hilbert space to include distributions (such as the Dirac delta ``function'') by using the rigged Hilbert space, namely the Gelfand triple. In the end, from a given plane-wave S-matrix, one can obtain a physically measurable probability only by convoluting it with wave packets.
}
Due to this finiteness of the Gaussian $S$-matrix, the probability for the transition $\Ket{\tx{in};\Pi_0}\HprPh\to\Ket{\tx{out};\Pi_1,\Pi_2}\Hprpp$ is simply its square: $\ab{S}^2$.\footnote{
\red{So far, we have not considered any boundary effect as we assume here that the interactions are negligible at $T_\tx{in}$ and $T_\tx{out}$; see also footnote~\ref{cannot dump interaction}.}
}
There is no need of the hand-waving argument of the momentum delta function $\delta^4\fn{0}$ becoming spacetime volume etc.

Using Eq.~\eqref{in and out states for scalar decay}, we get
\al{
S
	\approx
		\SBlpp\Bra{\Pi_1,\Pi_2}\h U\fn{T_\tx{out},T_\tx{in}}\Ket{\Pi_0}\SBrPh.
}
At the first order in the Dyson series~\eqref{Dyson series},
\al{
\h U\fn{T_\tx{out},T_\tx{in}}
	&=
		1-i\int_{T_\tx{in}}^{T_\tx{out}}\df t\int\df^3\bs x\,\h{\mc H}^\Ip_\tx{int}\fn{x}
		+\cdots,
}
the $S$-matrix becomes
\al{
S	&=
		i\int_{T_\tx{in}}^{T_\tx{out}}\df t\int\df^3\bs x\,
		\SBlpp\Bra{\Pi_1,\Pi_2}
			\h{\mc L}_\tx{int}^\Ip\fn{x}\Ket{\Pi_0}\SBrPh\nn
	&=	{i\kappa\ov2}\int_{T_\tx{in}}^{T_\tx{out}}\df t\int\df^3\bs x\,
		\SBlpp\Bra{\Pi_1,\Pi_2}\h\phi^\Ip\fn{x}\h\phi^\Ip\fn{x}\Ket{0}\,
		\Bra{0}\h\Phi^\Ip\fn{x}\Ket{\Pi_0}\SBrPh\nn
	&=	{i\kappa\ov\sqrt{2}}\int_{T_\tx{in}}^{T_\tx{out}}\df t\int\df^3\bs x\,
		f_{\phi,\sigma_1;\Pi_1}^*\fn{x}\,
		f_{\phi,\sigma_2;\Pi_2}^*\fn{x}\,
		f_{\Phi,\sigma_0;\Pi_0}\fn{x}.
}

\section{Gaussian $S$-matrix: separation of bulk and boundary effects}
\label{S-matrix section}

\begin{figure}
\centering
 \includegraphics[clip,width=.62\textwidth]{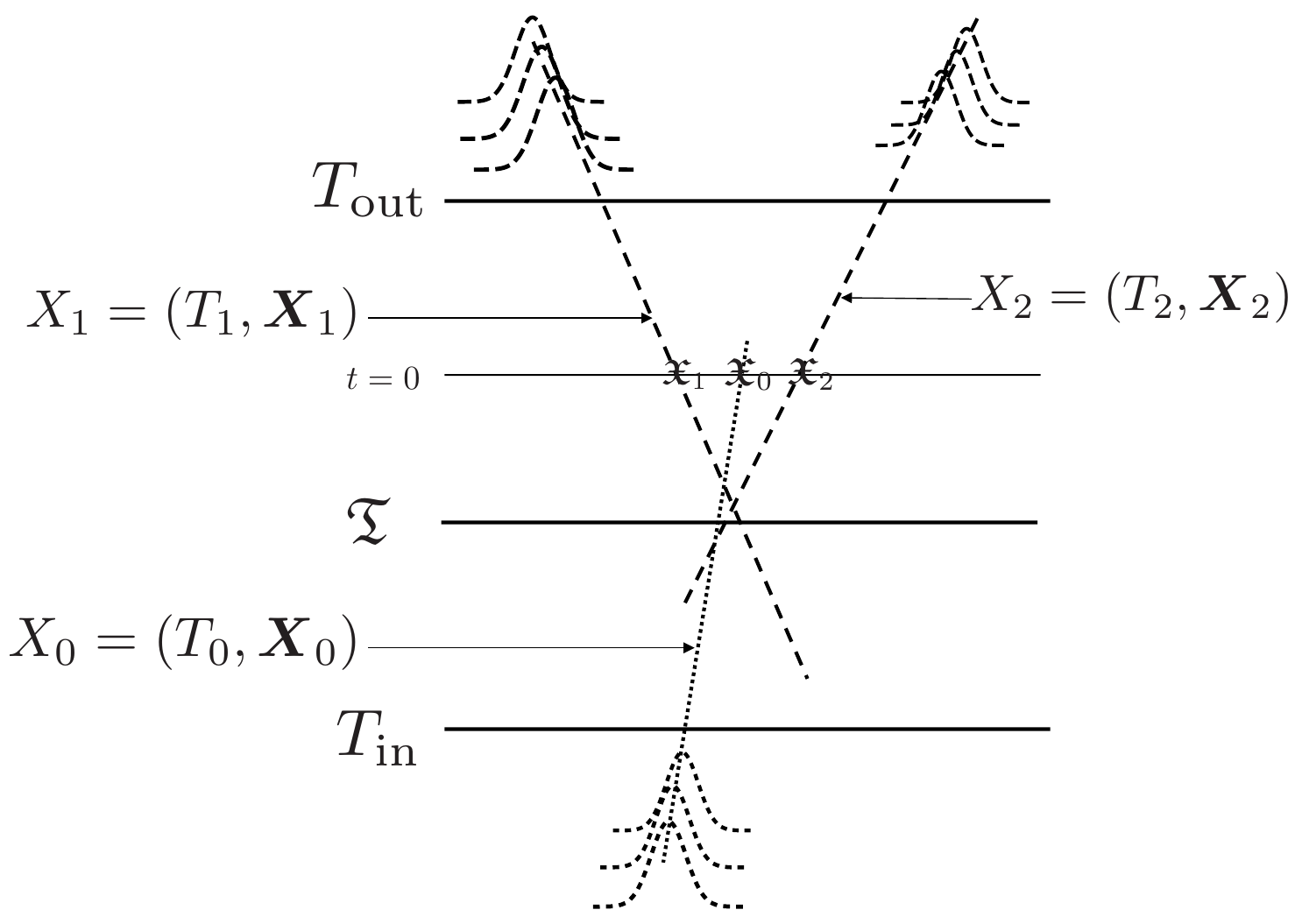}
 \caption{Schematic figure for a configuration with fixed $\Pi_A=\paren{\bs P_A,X_A}$ with $A=0,1,2$. ($\sigma_A$ are kept fixed throughout this paper.)
Each wave packet is defined at time~$T_A$ as a free Gaussian wave packet centered at $\bs X_A$. Within our leading saddle-point approximation, the widths of the wave packets do not change in time, see Eq.~\eqref{plane wave limit}, and therefore it does not really matter at which time each wave packet is set to be the free Gaussian wave packet. The wave packets intersect at the time~$\mf T$, around which the interactions occur most. $\bmf X_A$ is the location of the center of each wave packet at the (arbitrarily chosen) reference time $t=0$. At time $t$, the location of the center moves to $\bs\Xi_A\fn{t}=\bmf X_A+\bs V_At$.}
 \label{schematic figure}
\end{figure}

Now we compute the Gaussian $S$-matrix.
In Sec.~\ref{leading saddle point}, we obtain the $S$-matrix in the leading saddle-point approximation~\eqref{plane wave limit} for the large widths expansion.
In Sec.~\ref{spacetime integral}, we exactly integrate over the spacetime position $x$ of the interaction point.
In Sec.~\ref{separation section}, we separate the bulk and boundary effects.
In Sec.~\ref{large argument limit}, a limit of large argument is taken to get some physical insight.
A schematic figure for this section is presented in Fig.~\ref{schematic figure}.

\subsection{Saddle-point approximation in plane-wave limit}\label{leading saddle point}

With the leading saddle-point approximation~\eqref{plane wave limit} in the large width expansion for all the in and out wave packets, we obtain the $S$-matrix for a given configuration $\paren{\Pi_0,\Pi_1,\Pi_2}$:\footnote{
Recall that we abbreviate $\paren{\sigma_0,\Pi_0;\sigma_1,\Pi_1;\sigma_2,\Pi_2}$ to $\paren{\Pi_0,\Pi_1,\Pi_2}$.
}
\al{
S	&\to
		{i\kappa\ov\sqrt{2}}
		\paren{\prod_{A=0}^2\paren{\pi\sigma_A}^{-3/4}{1\ov\sqrt{2E_A}}}
			e^{-{\sigma_t\ov2}\paren{\delta\omega}^2-{\sigma_s\ov2}\paren{\delta\bs P}^2-{\mc R\ov2}+i\sqbr{\cdots}}\nn
	&\quad
			\times
			\int_{T_\tx{in}}^{T_\tx{out}}\df t\,e^{-{1\ov2\sigma_t}\sqbr{t-\paren{\mf T+i\sigma_t\delta\omega}}^2}
			\int\df^3\bs x\,e^{-{1\ov2\sigma_s}\sqbr{\bs x-\paren{\ol{\bmf X}+\ol{\bs V}t-i\sigma_s\delta\bs P}}^2},
				\label{plane wave limit of S}
}
where the symbols indicate the following:
\begin{itemize}
\item $E_A$ are the on-shell energies:
\al{
E_A	&:=	\sqrt{m_A^2+\bs P_A^2}&
	(A&=0,1,2),
	\label{on shell energy}
}
with $m_0:=m_\Phi$ and $m_a:=m_\phi$ ($a=1,2$) being their masses. (This is mere a rephrasing of Eq.~\eqref{on-shell energy}.)
\item $\bs V_A$ are the corresponding group velocities:
\al{
\bs V_A
	&:=	{\bs P_A\ov E_A}.
}
We may freely choose either variable $\bs P_A$ or $\bs V_A$, which are in one-to-one correspondence.
\item $\sqrt{\sigma_s}$ is the spatial size of the interaction region:
\al{
\sigma_s
	&:=	\paren{\sum_{A=0}^2{1\ov\sigma_A}}^{-1}.
		\label{sigma s defined}
}
Hereafter, we abbreviate e.g.\ $\sum_{A=0}^2$ to $\sum_A$. (We also let the lower-case letters $a,b,\dots$ run for the final states 1 and 2 such that $\sum_a:=\sum_{a=1}^2$, etc.)
\item The overline denotes the following weighted sum (and not the complex conjugate): For arbitrary scalar and three-vector quantities $C_A$ and $\bs Q_A$, respectively, we define
\al{
\ol C
	&:=	\sigma_s\sum_A{C_A\ov\sigma_A},	&
\ol{\bs Q}
	&:=	\sigma_s\sum_A{\bs Q_A\ov\sigma_A}.
		\label{bar defined}
}
We further define, for any $\bs Q_A$,
\al{
\Delta\bs Q^2
	&:=	\ol{\bs Q^2}-\ol{\bs Q}^2,
}
where $
\ol{\bs Q^2}
	=	\sigma_s\sum_A{\bs Q_A^2\ov\sigma_A}$ and
$\ol{\bs Q}^2
	=	\ol{\bs Q}\cdot\ol{\bs Q}
	=	\sigma_s^2\sum_{AB}{\bs Q_A\cdot\bs Q_B\ov\sigma_A\sigma_B}$, which follow from the definition~\eqref{bar defined}.
\item $\sqrt{\sigma_t}$ is the time-like size of the interaction region:
\al{
\sigma_t
	&:=	
		{\sigma_s\ov\Delta\bs V^2}.
		\label{sigma_t defined}
}
\item $\mf T$ is what we call the \emph{intersection time}, around which the interaction occurs:
\al{
\mf T
	&:=	\sigma_t{\ol{\bs V}\cdot\ol{\bmf X}-\ol{\bs V\cdot\bmf X}\ov\sigma_s}
	=	
		{\ol{\bs V}\cdot\ol{\bmf X}-\ol{\bs V\cdot\bmf X}\ov\Delta\bs V^2},
}
where $\bmf X_A=\bs\Xi_A\fn{0}$ is the location of the center of each wave packet at our reference time $t=0$:
\al{
\bmf X_A:=\bs X_A-\bs V_AT_A.
}
\item $\mc R$ is what we will call the \emph{overlap exponent} that gives the suppression factor accounting for the non-overlap of the wave packets at the intersection point:
\al{
\mc R
	&:=	{\Delta\bmf X^2\ov\sigma_s}-{\mf T^2\ov\sigma_t}
		.\label{definition of R}
} 
\item We define the mometum and energy shifts, etc:
\al{
\delta\bs P
	&:=	\bs P_1+\bs P_2-\bs P_0,	&
\delta E
	&:=	E_1+E_2-E_0,	&
\delta\omega
	&:=	\delta E-\ol{\bs V}\cdot\delta\bs P.
		\label{energy shift etc}
}
\item ``$i\sqbr{\cdots}$'' denotes the irrelevant pure imaginary terms that are independent of $x$.
We will neglect them hereafter as they disappear when we take the absolute square of $S$.
\end{itemize}
Note that each quantity defined in the above list is a fixed real number for a given configuration of the wave packets $\paren{\Pi_0,\Pi_1,\Pi_2}$.
Later we will treat $\bs X_a$ ($a=1,2$) as variables of six degrees of freedom; others $\mf T$, $\bs{\mf X}_a$, and $\mc R$ are dependent ones. (If we vary the final state momenta, then $\bs P_a$ ($a=1,2$) also become variables; others $\bs V_a$, $\sigma_t$, $\delta\bs P$, $\delta E$, and $\delta\omega$ become dependent ones accordingly.)

For any pair of three-vectors $\bs Q_A$ and $\bs Q_A'$ ($A=0,1,2$), we get
\al{
\ol{\bs Q\cdot\bs Q'}-\ol{\bs Q}\cdot\ol{\bs Q'}
	&=	\sigma_s^2\sqbr{
		{\delta\bs Q_1\cdot\delta\bs Q_1'\ov\sigma_0\sigma_1}
		+{\delta\bs Q_2\cdot\delta\bs Q_2'\ov\sigma_0\sigma_2}
		+{\paren{\delta\bs Q_1-\delta\bs Q_2}\cdot\paren{\delta\bs Q_1'-\delta\bs Q_2'}\ov\sigma_1\sigma_2}
		},
}
where we define, for any $\bs Q_A$,\footnote{
The abuse of notation for $\delta$ in Eq.~\eqref{energy shift etc} should be understood.
}
\al{
\delta\bs Q_a
	&:=	\bs Q_a-\bs Q_0.
		\label{delta Q defined}
}
Note that we always have $\delta\bs Q_1-\delta\bs Q_2=\bs Q_1-\bs Q_2$.
Especially,
\al{
\Delta\bs Q^2
	&=	\sigma_s^2\sqbr{
			{\paren{\delta\bs Q_1}^2\ov\sigma_0\sigma_1}
			+{\paren{\delta\bs Q_2}^2\ov\sigma_0\sigma_2}
			+{\paren{\delta\bs Q_1-\delta\bs Q_2}^2\ov\sigma_1\sigma_2}
			}\nn
	&=	\sigma_s^2\sqbr{
			{\paren{\bs Q_1-\bs Q_0}^2\ov\sigma_0\sigma_1}
			+{\paren{\bs Q_2-\bs Q_0}^2\ov\sigma_0\sigma_2}
			+{\paren{\bs Q_1-\bs Q_2}^2\ov\sigma_1\sigma_2}
			},
}
or more concretely,
\al{
\Delta\bmf X^2
	&=	\sigma_s^2
		\sqbr{
			{\paren{\delta\bs{\mf X}_1}^2\ov\sigma_0\sigma_1}
			+{\paren{\delta\bs{\mf X}_2}^2\ov\sigma_0\sigma_2}
			+{\paren{\delta\bs{\mf X}_1-\delta\bs{\mf X}_2}^2\ov\sigma_1\sigma_2}
			},\\
\Delta\bs V^2
	&=	\sigma_s^2
		\sqbr{
			{\paren{\delta\bs V_1}^2\ov\sigma_0\sigma_1}
			+{\paren{\delta\bs V_2}^2\ov\sigma_0\sigma_2}
			+{\paren{\delta\bs V_1-\delta\bs V_2}^2\ov\sigma_1\sigma_2}
			}.
				\label{Delta V squared}
}
Then we get
\al{
\sigma_t
	&=	{1\ov \sigma_s}\sqbr{
			{\paren{\delta\bs V_1}^2\ov\sigma_0\sigma_1}
			+{\paren{\delta\bs V_2}^2\ov\sigma_0\sigma_2}
			+{\paren{\delta\bs V_1-\delta\bs V_2}^2\ov\sigma_1\sigma_2}
				}^{-1},\\
\mf T
	&=	-\sigma_s\sigma_t\sqbr{
		{\delta\bs{\mf X}_1\cdot\delta\bs V_1\ov\sigma_0\sigma_1}
		+{\delta\bs{\mf X}_2\cdot\delta\bs V_2\ov\sigma_0\sigma_2}
		+{\paren{\delta\bs{\mf X}_1-\delta\bs{\mf X}_2}\cdot\paren{\delta\bs V_1-\delta\bs V_2}\ov\sigma_1\sigma_2}
		},\label{explicit T}\\
\mc R
	&=	\sigma_s\Bigg\{
			{\paren{\delta\bs{\mf X}_1}^2\ov\sigma_0\sigma_1}
			+{\paren{\delta\bs{\mf X}_2}^2\ov\sigma_0\sigma_2}
			+{\paren{\delta\bs{\mf X}_1-\delta\bs{\mf X}_2}^2\ov\sigma_1\sigma_2}
				\nn
	&\qquad
		-{\sigma_s\sigma_t}
			\sqbr{			
		{\delta\bs{\mf X}_1\cdot\delta\bs V_1\ov\sigma_0\sigma_1}
		+{\delta\bs{\mf X}_2\cdot\delta\bs V_2\ov\sigma_0\sigma_2}
		+{\paren{\delta\bs{\mf X}_1-\delta\bs{\mf X}_2}\cdot\paren{\delta\bs V_1-\delta\bs V_2}\ov\sigma_1\sigma_2}
				}^2\Bigg\}.\label{explicit R}
}
Note that for a parent particle at rest, $\bs P_0=0$, we may simply replace $\delta\bs V_a\to \bs V_a$. 
Expressions in various limits are shown in Appendix~\ref{limits section}.

Let us prove the non-negativity of $\mc R$.
In general, the weighted average for any real vector~$\bs Q$ satisfies
\al{
\Delta\bs Q^2
	=	\ol{\bs Q^2}-\ol{\bs Q}^2
	=	\ol{\paren{\bs Q-\ol{\bs Q}}^2}
	&\geq0.
}
From this, one can deduce the non-negativity of $\mc R$ as follows:
At time $t$, the center of each wave packet is located at
\al{
\bs Z_A
	&:=	\bmf X_A+\bs V_At.
}
The square completion of $\Delta\bs Z^2=\ol{\bs Z^2}-\ol{\bs Z}^2$ with respect to $t$ shows that $\Delta\bs Z^2$ takes its minimum value $\sigma_s\mc R$ at $t=\mf T$:
\al{
\Delta\bs Z^2
	&=	\Delta\bs V^2\paren{t-\mf T}^2+\sigma_s\mc R.
		\label{Delta Z equation}
}
As $\Delta\bs Z^2\geq0$ for any $t$, we obtain $\sigma_s\mc R\geq0$, hence the non-negativity of $\mc R$. 

In particular, if the center of all the three wave packets coincide at $\bs Z_A=\bs x$ at some time~$t$, then $\ol{\bs Z}=\bs x$ and $\Delta\bs Z^2=0$. Eq.~\eqref{Delta Z equation} shows that this can be the case when and only when $t=\mf T$ (for $\Delta\bs V^2>0$) and that we get no suppression in such a case, $\mc R=0$.

Let us see the physical meaning of $\mf T$.
Suppose that we recklessly take the particle limit $\sigma_t,\sigma_s\to0$ in the second line in Eq.~\eqref{plane wave limit of S} even though the expression itself is obtained in the contrary plane-wave expansion. 
Then we see that the interaction indeed occurs around the spacetime point
\al{
x=\paren{t,\bs x}\sim\paren{\mf T,\ol{\bmf X}+\ol{\bs V}\mf T},
	\label{intersection point}
}
which we call the \emph{intersection point}.

One can show (without taking the particle limit) that the intersection point~\eqref{intersection point} is transformed properly by the spacetime translation: By a constant spacetime translation
\al{
\bs X_A
	&\to	\bs X_A+\bs d,\nn
T_A	&\to	T_A+d^0,
}
the center of each wave packet (at $t=0$) and its average transform as
\al{
\bmf X_A
	&\to	\bmf X_A+\bs d-\bs V_Ad^0,\\
\ol{\bmf X}
	&\to	\ol{\bmf X}+\bs d-\ol{\bs V}d^0,
}
and hence
\al{
\mf T
	&\to	\mf T+d^0,\\
\paren{\ol{\bmf X}+\ol V\mf T}
	&\to	\paren{\ol{\bmf X}+\ol V\mf T}+\bs d.
}
One can also check that the overlap exponent $\mc R$ is translationally invariant (as it should physically be):
\al{
\mc R
	&\to	\mc R.
}

In particular, we may choose
\al{
\bs d=\bs V_0d^0,
	\label{particular translation}
}
such that the center of the initial wave packet at $t=0$, $\bs{\mf X}_0=\bs X_0-\bs V_0T_0$, is kept invariant.
Then the center of each final-state wave packet, $\bmf X_a$, is shifted as\footnote{
The average over the initial and final states is shifted as
$\ol{\bs{\mf X}}
	\to	\ol{\bs{\mf X}}-\paren{\ol{\bs V}-\bs V_0}d^0$.
}
\al{
\bs{\mf X}_a
	&\to	\bs{\mf X}_a-\delta\bs V_ad^0,&
\delta\bs{\mf X}_a
	&\to	\delta\bs{\mf X}_a-\delta\bs V_ad^0.
			\label{transformation of german X}
}
Later, this translation will correspond to the zero mode~\eqref{zero eigenvector given}.

\subsection{Spacetime integral over position of interaction point}\label{spacetime integral}
One can exactly perform the Gaussian integrals over the interaction point $x=\paren{t,\bs x}$ in Eq.~\eqref{plane wave limit of S} to get
\al{
S	&=	{i\kappa\ov\sqrt2}\paren{\prod_A\paren{\pi\sigma_A}^{-3/4}{1\ov\sqrt{2E_A}}}
		e^{-{\sigma_t\ov2}\paren{\delta\omega}^2-{\sigma_s\ov2}\paren{\delta\bs P}^2-{\mc R\ov2}}
		\paren{2\pi\sigma_s}^{3/2}\sqrt{2\pi\sigma_t}\,G\fn{\mf T},
			\label{S Gauss-integrated}
}
where we have defined the \emph{window function}:
\al{
G\fn{\mf T}
	&:=	\int_{T_\tx{in}}^{T_\tx{out}}{\df t\ov\sqrt{2\pi\sigma_t}}e^{-{1\ov2\sigma_t}\paren{t-\mf T-i\sigma_t\delta\omega}^2}\nn
	&=	{1\ov2}\sqbr{
			\erf\fn{\mf T-T_\tx{in}+i\sigma_t\delta\omega\ov\sqrt{2\sigma_t}}
			-\erf\fn{\mf T-T_\tx{out}+i\sigma_t\delta\omega\ov\sqrt{2\sigma_t}}
			},
			\label{window function defined}
}
in which 
\al{
\erf\fn{z}
	&:=	{2\ov\sqrt{\pi}}\int_0^ze^{-x^2}\df x
}
is the Gauss error function.
In the small and large $\ab{z}$ limits, its (asymptotic) expansion reads, respectively,
\al{
\erf\fn{z}
	&=	{2z\ov\sqrt{\pi}}+\Or{z^3},\\
\erf\fn{z}
	&=	\sgn\fn{z}+e^{-z^2}\paren{-{1\ov\sqrt{\pi}z}+\Or{z^{-3}}},
}
where we have defined a sign function for a complex variable:
\al{
\sgn\fn{z}
	&:=	\begin{cases}
		1	&	\tx{for $\Re z>0$ or ($\Re z=0$ and $\Im z>0$),}\\
		-1	&	\tx{for $\Re z<0$ or ($\Re z=0$ and $\Im z<0$),}\\
		0	&	\tx{for $z=0$.}
		\end{cases}
}

From Eq.~\eqref{S Gauss-integrated}, we see that the $S$-matrix is exponentially suppressed unless the momentum is nearly conserved, $\delta\bs P\sim0$. This is also the case for the energy conservation $\delta\omega\sim0$ except in the boundary regions, at which the translational invariance is explicitly broken; see Sec.~\ref{large argument limit} below. As said above, the overlap exponent $\mc R$ gives another suppression when the wave packets do not overlap.

\subsection{Separation of bulk and boundary effects}\label{separation section}

\begin{figure}
\centering
 \includegraphics[clip,width=.45\textwidth]{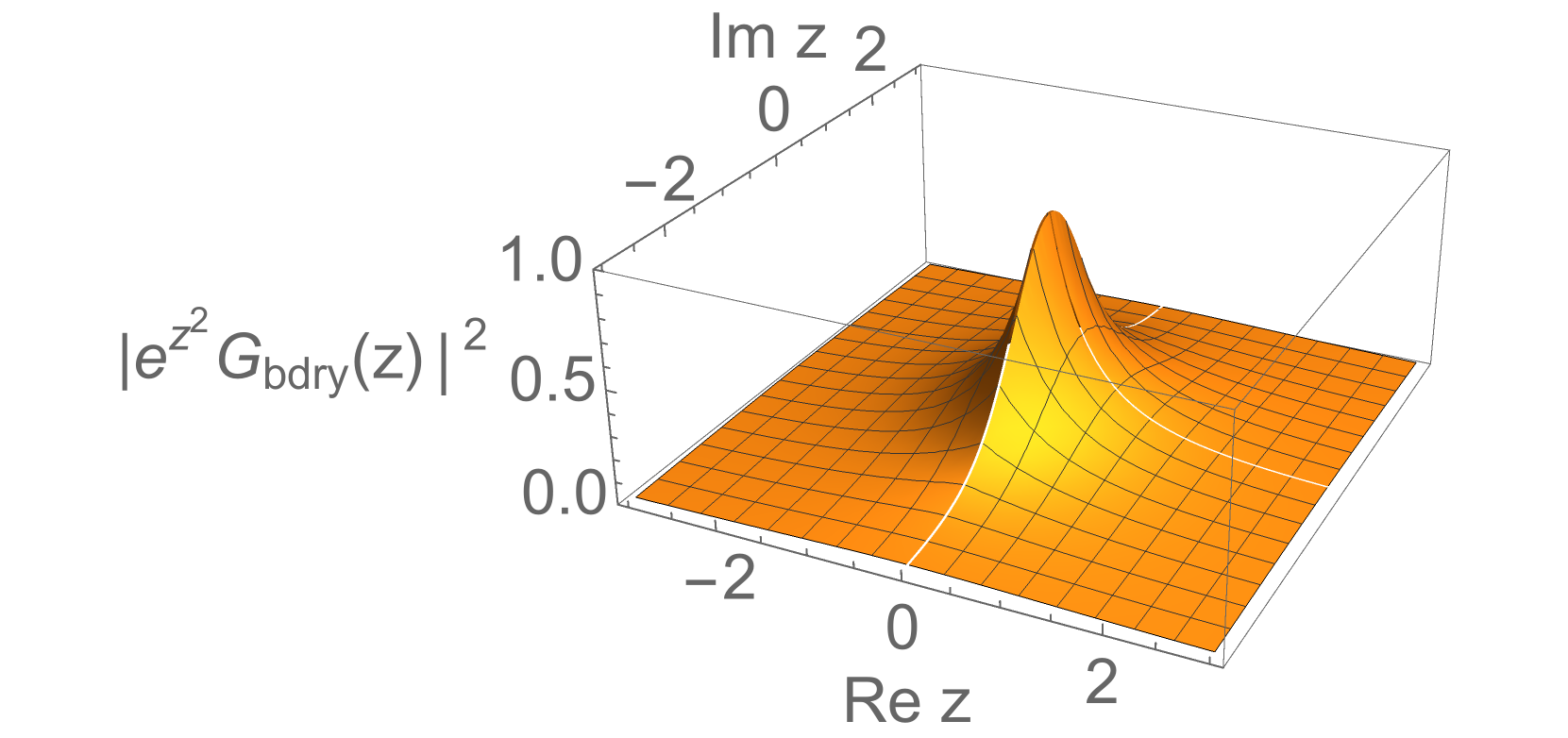}
 \includegraphics[clip,width=.49\textwidth]{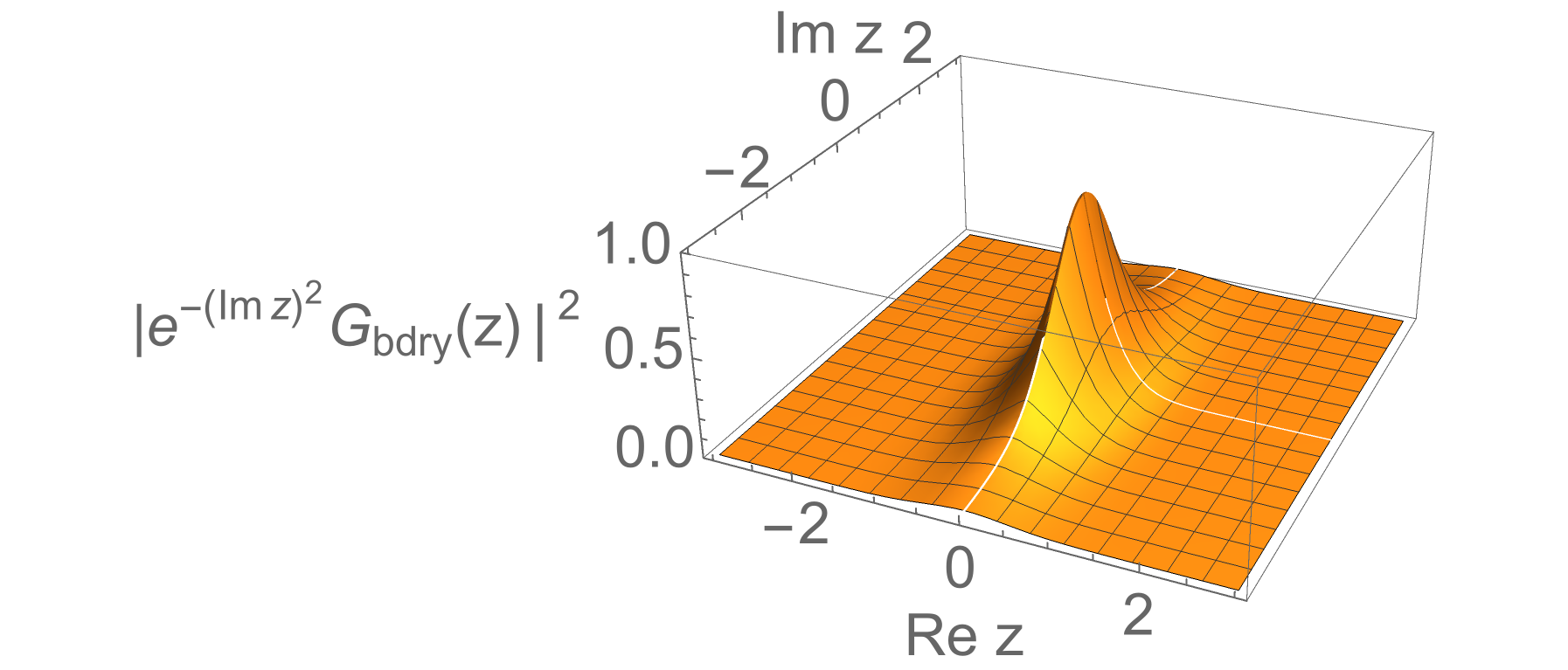}
 \caption{Normalized boundary function $\ab{e^{z^2}G_\tx{bdry}\fn{z}}^2$ (left) and the combination that appears in physical setup $\ab{e^{-\paren{\Im z}^2}G_\tx{bdry}\fn{z}}^2$ (right).
 }
 \label{boundary fig}
\end{figure}

It is convenient to separate the window function~\eqref{window function defined} into the bulk part and the in- and out-boundary ones:
\al{
G\fn{\mf T}
	&=	G_\tx{bulk}\fn{\mf T}
		+G_\tx{in-bdry}\fn{\mf T}
		+G_\tx{out-bdry}\fn{\mf T},
			\label{G separated}
}
where
\al{
G_\tx{bulk}\fn{\mf T}
	&:=	{1\ov2}\sqbr{
			\sgn\fn{\mf T-T_\tx{in}+i\sigma_t\delta\omega\ov\sqrt{2\sigma_t}}
			-\sgn\fn{\mf T-T_\tx{out}+i\sigma_t\delta\omega\ov\sqrt{2\sigma_t}}},
				\label{G bulk}\\
G_\tx{in-bdry}\fn{\mf T}
	&:=	{1\ov2}\sqbr{
			\erf\fn{\mf T-T_\tx{in}+i\sigma_t\delta\omega\ov\sqrt{2\sigma_t}}
			-\sgn\fn{\mf T-T_\tx{in}+i\sigma_t\delta\omega\ov\sqrt{2\sigma_t}}},\nn
G_\tx{out-bdry}\fn{\mf T}
	&:=	{1\ov2}\sqbr{
			\sgn\fn{\mf T-T_\tx{out}+i\sigma_t\delta\omega\ov\sqrt{2\sigma_t}}
			-\erf\fn{\mf T-T_\tx{out}+i\sigma_t\delta\omega\ov\sqrt{2\sigma_t}}
			}.
		\label{G boundary}
}
One can rewrite the boundary parts:
\al{
G_\tx{in-bdry}\fn{\mf T}
	&=	{1\ov2}G_\tx{bdry}\fn{\mf T-T_\tx{in}+i\sigma_t\delta\omega\ov\sqrt{2\sigma_t}},\\
G_\tx{out-bdry}\fn{\mf T}
	&=	-{1\ov2}G_\tx{bdry}\fn{\mf T-T_\tx{out}+i\sigma_t\delta\omega\ov\sqrt{2\sigma_t}},
}
where
\al{
G_\tx{bdry}\fn{z}
	&:=	\erf\fn{z}-\sgn\fn{z}.
}
More explicitly, the bulk part reads
\al{
G_\tx{bulk}\fn{\mf T}
	&=	\begin{cases}
		1	&	(T_\tx{in}<\mf T<T_\tx{out}),\\
		0	&	(\mf T<T_\tx{in}\tx{ or }T_\tx{out}<\mf T),\\
		\theta\fn{\delta\omega}	&	(\mf T=T_\tx{in}),\\
		\theta\fn{-\delta\omega}	&	(\mf T=T_\tx{out}),
		\end{cases}
			\label{explicit bulk G}
}
where
\al{
\theta\fn{x}
	&=	{1+\sgn\fn{x}\ov2}
	=	\begin{cases}
		1	&	(x>0),\\
		{1\ov2}
			&	(x=0),\\
		0	&	(x<0),
		\end{cases}
}
is the step function.\footnote{
As we see in Eq.~\eqref{explicit bulk G}, this step function appears only at $\mf T=T_\tx{in/out}$ and hence does not contribute when summed with $G_\tx{bdry}$ and integrated over $\mf T$.
That is, it appears only at $\Re z=0$ and does not contribute when integrated over $\Re z$ in Fig.~\ref{boundary fig}. This might be non-vanishing for a more realistic non-Gaussian wave packet.
}

We note that $G_\tx{bdry}\fn{z}$ is discontinuous at $\Re z=0$ but the combination $\ab{e^{z^2}G_\tx{bdry}\fn{z}}^2$ is continuous and finite everywhere on the complex $z$ plane (except at the origin $z=0$); see Fig.~\ref{boundary fig}. Especially in the limit $\ab{z}\to\infty$, we obtain\footnote{\label{realistic combination}
In terms of the relevant combination, we get
\als{
e^{-2\paren{\Im z}^2}\ab{G_\tx{bdry}\fn{z}}^2
	&\to	{e^{-2\paren{\Re z}^2}\ov\pi\ab{z}^2}.
}
}
\al{
\ab{e^{z^2}G_\tx{bdry}\fn{z}}^2
	&\to	{1\ov\pi\ab{z}^2}.
}

\begin{figure}
\centering
 \includegraphics[clip,width=.5\textwidth]{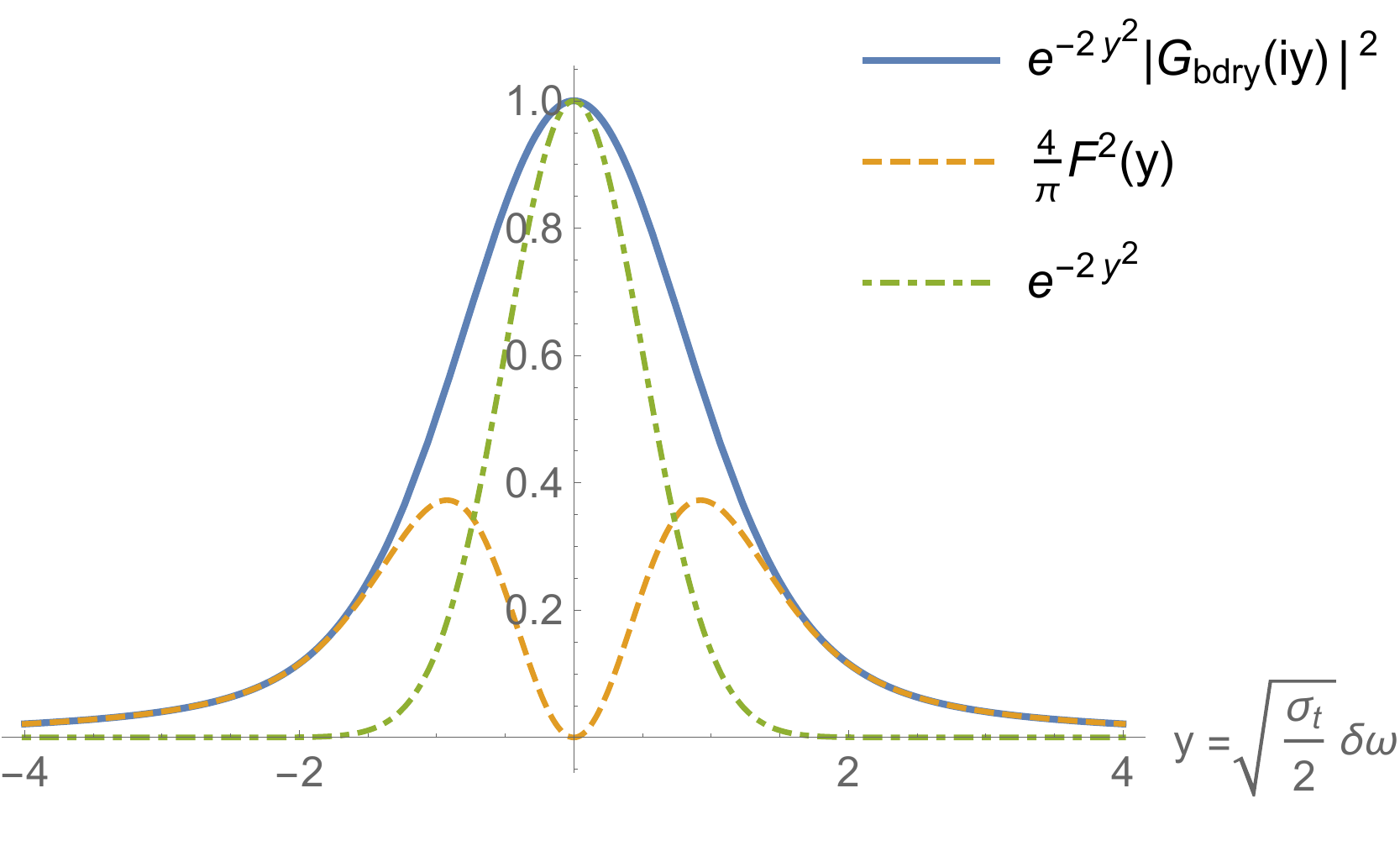}
 \caption{
 We plot $e^{-2y^2}\ab{G_\tx{bdry}\fn{iy}}^2$ (solid), ${4\ov\pi}F^2\fn{y}$ (dashed), and $e^{-2y^2}$ (dot-dashed) as a function of $y=\sqrt{\sigma_t\ov2}\delta\omega$, which corresponds to right on either boundary $\mf T=T_\tx{in/out}$. The solid line is the sum of the dashed and dot-dashed lines and corresponds to the ridge line at $\Re z=0$ in Fig.~\ref{boundary fig}.
 }
 \label{imaginary fig}
\end{figure}

The explicit formula in the boundary limit $\ab{\mf T-T_\tx{in/out}}\ll\sigma_t\delta\omega$ is
\al{
G_\tx{bdry}\fn{\mf T-T_\tx{in/out}+i\sigma_t\delta\omega\ov\sqrt{2\sigma_t}}
	&\to
		\pm{2i\ov\sqrt{\pi}}F\fn{\sqrt{\sigma_t\ov2}\delta\omega}
		e^{{\sigma_t\ov2}\paren{\delta\omega}^2}
		\mp
		\begin{cases}
		\sgn\fn{\delta\omega}
			&	\tx{for }\mf T=T_\tx{in/out},\\
		\sgn\fn{\mf T-T_\tx{in/out}}
			&	\tx{for }\mf T\neq T_\tx{in/out},
		\end{cases}
}
that is,\footnote{
We have assumed $\mf T-T_\tx{in/out}+i\sigma_t\delta\omega\neq0$ in writing $\sgn^2=1$.
}
\al{
e^{-\sigma_t\paren{\delta\omega}^2}\ab{G_\tx{bdry}\fn{\mf T-T_\tx{in/out}+i\sigma_t\delta\omega\ov\sqrt{2\sigma_t}}}^2
	&\to	
		{4\ov\pi}F^2\fn{\sqrt{\sigma_t\ov2}\delta\omega}+e^{-\sigma_t\paren{\delta\omega}^2},
		\label{Dawson}
}
where
\al{
F\fn{x}
	&:=	e^{-x^2}\int_0^xe^{z^2}\df z
	=	-i{\sqrt{\pi}\ov2}e^{-x^2}\erf\fn{ix}
}
is the Dawson function, whose (asymptotic) expansions read
\al{
F\fn{x}
	&=	x+\Or{x^3},\\
F\fn{x}
	&=	-i{\sqrt{\pi}\ov2}e^{-x^2}\sgn\fn{ix}
		+{1\ov2x}+\Or{1\ov x^3}.
}
More explicitly, the large $\sqrt{\sigma_t}\delta\omega$ expansion gives
\al{
{4\ov\pi}F^2\fn{\sqrt{\sigma_t\ov2}\delta\omega}+e^{-\sigma_t\paren{\delta\omega}^2}
	&=	{2\ov\pi}{1\ov\sigma_t\paren{\delta\omega}^2}+\Or{{1\ov\sigma_t^2\paren{\delta\omega}^4}}.
		\label{Dawson expanded}
}
In Fig.~\ref{imaginary fig}, we plot $G_\tx{bdry}$ right at either boundary $\mf T=T_\tx{in/out}$.

\subsection{Limit of large argument}\label{large argument limit}
In the limit
\al{
{\ab{\mf T-T_\tx{in}+i\sigma_t\delta\omega}\ov\sqrt{\sigma_t}},
{\ab{\mf T-T_\tx{out}+i\sigma_t\delta\omega}\ov\sqrt{\sigma_t}}\gg1,
}
the possible leading contributions are
\al{
G\fn{\mf T}
	&\to	{1\ov2}\bigg[
			\sgn\fn{\mf T-T_\tx{in}+i\sigma_t\delta\omega\ov\sqrt{2\sigma_t}}
			-\sgn\fn{\mf T-T_\tx{out}+i\sigma_t\delta\omega\ov\sqrt{2\sigma_t}}
			\nn
	&\phantom{\to{1\ov2}\bigg[}
		-e^{-{\paren{\mf T-T_\tx{in}}^2\ov2\sigma_t}+{\sigma_t\ov2}\paren{\delta\omega}^2-i\,\delta\omega\paren{\mf T-T_\tx{in}}}\sqrt{2\sigma_t\ov\pi}{1\ov \mf T-T_\tx{in}+i\sigma_t\delta\omega}\nn
	&\phantom{\to{1\ov2}\bigg[}
		+e^{-{\paren{\mf T-T_\tx{out}}^2\ov2\sigma_t}+{\sigma_t\ov2}\paren{\delta\omega}^2-i\,\delta\omega\paren{\mf T-T_\tx{out}}}\sqrt{2\sigma_t\ov\pi}{1\ov \mf T-T_\tx{out}+i\sigma_t\delta\omega}
		\bigg].
		\label{limit of G}
}
We see that the range of $\mf T$ in this limit can be separated into the following regions:
\begin{itemize}
\item In the \emph{bulk region}
\al{
\ab{\mf T-T_\tx{in}},\ab{\mf T-T_\tx{out}}\gg\sigma_t\delta\omega,
	\label{bulk region}
}
where the intersection time $\mf T$ is well separated from both the boundary times $T_\tx{in}$ and $T_\tx{out}$, we obtain
\al{
G\fn{\mf T}
	&\to
		W\fn{\mf T}
	:=	\begin{cases}
			1	&	(T_\tx{in}<\mf T<T_\tx{out}),\\
			0	&	\tx{(otherwise),}
			\end{cases}
			\label{bulk region G}
}
hence the name ``window function.''
\item In the \emph{in} and \emph{out boundary regions}
$\mf T\sim T_\tx{in}$ and $T_\tx{out}$ (namely $\ab{\mf T-T_\tx{in}}\lesssim\sigma_t\delta\omega$ and $\ab{\mf T-T_\tx{out}}\lesssim\sigma_t\delta\omega$), the contribution from the second and third lines, respectively, in Eq.~\eqref{limit of G} becomes sizable:
\al{
e^{-\sigma_t\paren{\delta\omega}^2}\ab{G\fn{\mf T}}^2
	&\to
		e^{-{\paren{\mf T-T_\tx{in/out}}^2\ov\sigma_t}}{2\ov\pi}{1\ov{\paren{\mf T-T_\tx{in/out}}^2\ov\sigma_t}+\sigma_t\paren{\delta\omega}^2}.
			\label{G limit}
}
We see that the exponential suppression $e^{-\sigma_t\paren{\delta\omega}^2}$ for $\sigma_t\paren{\delta\omega}^2\gg1$ becomes absent in the boundary region.
\end{itemize}
In Refs.~\cite{Ishikawa:2014eoa,Ishikawa:2013kba,Ishikawa:2014uma,Ishikawa:2016lnn}, the authors have claimed that contributions from the boundary region can become non-negligible and that there can be physical consequences.\footnote{\label{on boundary effects}
One might need a justification of placing the interaction around $T_\tx{in/out}$ that are defined to be the times at which the very interactions are negligible; see Eq.~\eqref{in and out states for scalar decay}. Note that this contradiction \red{in identifying the in-state with the free state at the remote past, as in Eq.~\eqref{in wave packet},} already exists in the ordinary plane-wave computation of the \emph{decay} \red{rate} because the interaction for the decay never becomes negligible even in the infinite past limit\red{, and one needs to dump the interaction by hand by introducing the $\pm i\epsilon$ \blue{term}}.
Better treatment would be to take into account the production process of the parent particle, namely, to compute the $\phi\phi\to\phi\phi$ scattering and the one-loop correction to it in the wave-packet formalism, which will be presented elsewhere.
}
In this paper, we leave this issue open and proceed by taking into account only the bulk region contribution~\eqref{bulk region G}; we will briefly comment on the boundary effects in Sec.~\ref{boundary comment section}.

Eq.~\eqref{G limit} appears singular in the simultaneous limit $\ab{\mf T-T_\tx{in/out}}\to0$ and $\sigma_t\delta\omega\to0$.
This apparent singularity is an artifact of first taking the limit~\eqref{limit of G}: 
If we take the limit $\ab{\mf T-T_\tx{in/out}+i\sigma_t\delta\omega}\ll\sqrt{\sigma_t}$ in the original expression~\eqref{window function defined}, we obtain
\al{
e^{-\sigma_t\paren{\delta\omega}^2}\ab{G\fn{\mf T}}^2
	&\to	{\paren{\mf T-T_\tx{in/out}}^2+\paren{\sigma_t\delta\omega}^2\ov2\pi\sigma_t}.
		\label{opposite limit}
}
It is manifest that we have no singularity.


\section{Decay probability: derivation of Fermi's golden rule}\label{decay probability section}
Recalling the (over-)completeness of the Gaussian basis~\eqref{wave packet completeness}, we see that the decay probability into an infinitesimal phase-space range $[\bs X_a,\bs X_a+\df\bs X_a]$ and $[\bs P_a,\bs P_a+\df\bs P_a]$ ($a=1,2$) is
\al{
\df P	
	&=	{\df^3\bs X_1\,\df^3\bs P_1\ov\paren{2\pi}^3}{\df^3\bs X_2\,\df^3\bs P_2\ov\paren{2\pi}^3}\ab{S}^2\nn
	&=	
		{\kappa^2\ov2}
		{1\ov2E_0}
		{\df^3\bs P_1\ov\paren{2\pi}^{3}2E_1}
		{\df^3\bs P_2\ov\paren{2\pi}^{3}2E_2}
		\paren{2\pi}^4
		\paren{\sqrt{\sigma_t\ov\pi}e^{-\sigma_t\paren{\delta\omega}^2}}
		\paren{\paren{\sigma_s\ov\pi}^{3/2}e^{-\sigma_s\paren{\delta\bs P}^2}}
			\nn
	&\qquad\times
		\sqrt{
		{\sigma_t\ov\pi^5}
		\paren{\sigma_s\ov\sigma_0\sigma_1\sigma_2}^3}\,
		\df^3\bs X_1\,\df^3\bs X_2\,
		e^{-\mc R}\,\ab{G\fn{\mf T}}^2.
	\label{dP full}
}
We note that this expression is exact up to the leading saddle point approximation~\eqref{plane wave limit}.

In Sec.~\ref{diagonalization section}, we show how to diagonalize the overlap exponent $\mc R$.
In Sec.~\ref{bulk section}, we focus on the bulk contribution $G_\tx{bulk}$ and derive the Fermi's golden rule.
In Sec.~\ref{boundary comment section}, we briefly comment on the boundary contribution $G_\tx{bdry}$.

\subsection{Diagonalization of overlap exponent}\label{diagonalization section}

Now we want to perform the Gaussian integral over the central positions of the wave packets~$\bs X_a$.
We may rewrite $\mc R$, in the matrix notation, as follows:
\al{
\mc R
	&=	\bmat{\delta\bs{\mf X}_1^\t&
				\delta\bs{\mf X}_2^\t}
				\mc M
			\bmat{\delta\bs{\mf X}_1\\ \delta\bs{\mf X}_2},
}
where the superscript ``$\t$'' denotes the transposition;
as defined in Eq.~\eqref{delta Q defined},
\al{
\delta\bs{\mf X}_a
	&:=	\bs{\mf X}_a-\bs{\mf X}_0 &
\delta\bs V_a
	&:=	\bs V_a-\bs V_0,
}
for $a=1,2$; and
$\mc M$ is the following real symmetric $6 \times 6$ matrix:
\al{
\mc M
	&=	{\sigma_s\ov\sigma_1\sigma_2}\bmat{1&-1\\ -1&1}
		+{\sigma_s\ov\sigma_0}\bmat{{1\ov\sigma_1}&0\\ 0&{1\ov\sigma_2}}
		+\sigma_s^2\sigma_t\bmat{
			\wt{\delta\bs V_1}\,\wt{\delta\bs V_1}^\t&
			\wt{\delta\bs V_1}\,\wt{\delta\bs V_2}^\t\\
			\wt{\delta\bs V_2}\,\wt{\delta\bs V_1}^\t&
			\wt{\delta\bs V_2}\,\wt{\delta\bs V_2}^\t
			},
}
in which
\al{
\wt{\delta\bs V_1}
	&:=	{\delta\bs V_1\ov\sigma_0\sigma_1}+{\delta\bs V_1-\delta\bs V_2\ov\sigma_1\sigma_2}
	=	{1\ov\sigma_1}\sqbr{
			\paren{{1\ov\sigma_0}+{1\ov\sigma_2}}\delta\bs V_1
			-{\delta\bs V_2\ov\sigma_2}
			},	\\
\wt{\delta\bs V_2}
	&:=	{\delta\bs V_2\ov\sigma_0\sigma_2}-{\delta\bs V_1-\delta\bs V_2\ov\sigma_1\sigma_2}
	=	{1\ov\sigma_2}\sqbr{
			\paren{{1\ov\sigma_0}+{1\ov\sigma_1}}\delta\bs V_2
			-{\delta\bs V_1\ov\sigma_1}}.
}
Hereafter, we employ the shifted $\delta\bmf X_a=\bs X_a-\paren{\bs V_aT_a+\bs X_0-\bs V_0T_0}$ as six integration variables.

One can check that $\mc M$ has a zero eigenvector:
\al{
\mc M\ora{\mc X_0}
	&=	0,	&
\ora{\mc X_0}
	&=	{1\ov\sqrt{\paren{\delta\bs V_1}^2+\paren{\delta\bs V_2}^2}}
		\bmat{\delta\bs V_1\\ \delta\bs V_2},
		\label{zero eigenvector given}
}
where we have normalized $\ora{\mc X_0}$ as $\ora{\mc X_0}^\t\,\ora{\mc X_0}=1$.
This is a direct consequence of the translational invariance under Eq.~\eqref{transformation of german X}.
This zero-mode will eventually give the factor~$T_\tx{out}-T_\tx{in}$, which is the characteristic of the Fermi's golden rule.

Writing other five normalized eigenvectors $\ora{\mc X_I}$ ($I=1,\dots,5$), we get\footnote{\label{zero eigenvector}
Recall that the zero eigenvector $\ora{\mc X_0}$ drops out of the spectral representation:
\als{
\mc M
	&=	\bmat{\ora{\mc X_0}&\ora{\mc X_1}&\cdots&\ora{\mc X_5}}
		\bmat{0&\\&\lambda_1&\\ &&\ddots\\&&&\lambda_5}
		\bmat{\ora{\mc X_0}^\t\\ \ora{\mc X_1}^\t\\ \vdots\\ \ora{\mc X_5}^\t}
	=	\bmat{\ora{0}&\ora{\mc X_1}&\cdots&\ora{\mc X_5}}
		\bmat{0&\\&\lambda_1&\\ &&\ddots\\&&&\lambda_5}
		\bmat{\ora{0}^\t\\ \ora{\mc X_1}^\t\\ \vdots\\ \ora{\mc X_5}^\t}.
}
}
\al{
\mc O^\t\mc M\mc O
	&=	\diag\paren{0,\lambda_1,\dots,\lambda_5},&
\mc M
	&=	\sum_{I=1}^5\lambda_I\ora{\mc X_I}\ora{\mc X_I}^\t,
}
where $\mc O:=\bmat{\ora{\mc X_0}&\ora{\mc X_1}&\cdots&\ora{\mc X_5}}$.
Explicit forms of the other five eigenvalues $\lambda_1,\dots,\lambda_5$ are
\al{
&{\sigma_s\ov2\sigma_0}\paren{{1\ov\sigma_1}+{1\ov\sigma_2}}
	+{\sigma_s\ov\sigma_1\sigma_2}\paren{1\pm\sqrt{1+\paren{\sigma_1-\sigma_2\ov2\sigma_0}^2}},&
&{\sigma_s\sigma_t\ov\sigma_0\sigma_1\sigma_2}
	\paren{\paren{\delta\bs V_1}^2+\paren{\delta\bs V_2}^2},
	\label{eigenvalues of M}
}
where we have assumed $\Delta\bs V^2>0$ in deriving the former, which is two-fold degenerate per each $\pm$ sign, providing four of the five.\footnote{
The eigenvector for the latter is proportional to
\als{
&\bmat{
	{\paren{\delta\bs V_1}^2+\paren{\delta\bs V_2}^2\ov2}\delta\bs V_2
	-\paren{	
			\paren{\delta\bs V_1\cdot\delta\bs V_2}
			-{\sigma_1-\sigma_2\ov2\sigma_0}\paren{\delta\bs V_2}^2}
			\delta\bs V_1\\
	{\paren{\delta\bs V_1}^2+\paren{\delta\bs V_2}^2\ov2}\delta\bs V_1
	-\paren{	
			\paren{\delta\bs V_1\cdot\delta\bs V_2}
			+{\sigma_1-\sigma_2\ov2\sigma_0}\paren{\delta\bs V_1}^2}
			\delta\bs V_2
}
	,
}
which can be explicitly checked to be orthogonal to the zero-eigenvector $\bmat{\delta\bs V_1\\ \delta\bs V_2}$.
}
After some computation, we obtain
\al{
\prod_{I=1}^5\lambda_I
	&=	\sigma_t\paren{\sigma_s\ov\sigma_0\sigma_1\sigma_2}^3
		\paren{\paren{\delta\bs V_1}^2+\paren{\delta\bs V_2}^2}.
			\label{prod eigenvalues}
}

We define new integration variables $\paren{y_0,y_1,\dots,y_5}$ by
\al{
\bmat{y_0\\ \vdots\\ y_5}
	&=	\mc O^\t\bmat{\delta\bs{\mf X}_1\\ \delta\bs{\mf X}_2},	&
\bmat{\delta\bs{\mf X}_1\\ \delta\bs{\mf X}_2}
	&=	\mc O\bmat{y_0\\ \vdots\\ y_5}.
		\label{new variables}
}
In particular, we get
\al{
y_0
	&=	{\delta\bs V_1\cdot\delta\bs{\mf X}_1
			+\delta\bs V_2\cdot\delta\bs{\mf X}_2
			\ov\sqrt{\paren{\delta\bs V_1}^2+\paren{\delta\bs V_2}^2}}
			.
}
As said above, the integral over $y_0$ does not have a Gaussian suppression and will yield the factor $\propto\paren{T_\tx{out}-T_\tx{in}}$.
Note that
\al{
\df^3\bs X_1\,\df^3\bs X_2=\df^6y
}
as $\mc O$ is \blue{a special} orthogonal matrix.

\subsection{Bulk contribution: derivation of Fermi's golden rule}\label{bulk section}
Now we concentrate on the bulk contribution~\eqref{G bulk}.
Physically, this takes into account the bulk region~\eqref{bulk region}, in which the window function takes the particularly simple form~\eqref{bulk region G}, by which the spatial $\df^6y$ integral is confined within the range that satisfies
\al{
T_\tx{in}<\mf T<T_\tx{out},
}
where the explicit form of $\mf T$ is given in Eq.~\eqref{explicit T}.
We note that $\mf T$ is linear in $\delta\bmf X_a$, and hence in $y_I$.

In a typical non-singular configuration of $\paren{\bs P_1,\bs P_2}$ with $\Delta\bs V^2>0$, the integral over all the other five variables $y_1,\dots,y_5$ are confined by the Gaussian factor within the range of the order of $\sqrt{\sigma_s}$; see Eq.~\eqref{eigenvalues of M}. By definition, the interaction point of the bulk region is well separated from the boundaries, and hence the window function can be regarded as unity for the integral over $y_1,\dots,y_5$.\footnote{\label{particle limit comment}
Though we have taken the leading saddle-point approximation in the large $\sigma_A$ expansion in obtaining Eq.~\eqref{plane wave limit of S}, we still consider that the wave packets are well localized compared to the whole spacetime volume in which the decay occurs, say, $\ab{\mf T-T_\tx{in/out}}\gg\sqrt{\sigma_s}$. This is consistent with the treatment of the current work restricted within the bulk region.
}
That is, we may safely perform each integral over these five variables as simply Gaussian:
\al{
\int\df^5y\,e^{-\mc R}
	&=	\sqrt{\pi^5\ov\prod_{I=1}^5\lambda_I}
	=	\sqrt{{\pi^5\ov\sigma_t}\paren{\sigma_0\sigma_1\sigma_2\ov\sigma_s}^3}
		{1\ov\sqrt{\paren{\delta\bs V_1}^2+\paren{\delta\bs V_2}^2}},
		\label{five dimensional integral}
}
in which we used the product of eigenvalues given in Eq.~\eqref{prod eigenvalues}.

Rewriting $\delta\bmf X_a$ in Eq.~\eqref{explicit T} by $y_0,\dots,y_5$ using the latter of Eq.~\eqref{new variables}, $\bmat{\delta\bmf X_1\\ \delta\bmf X_2}=\sum_{I=0}^5\ora{\mc X_I}\,y_I$, we can read off the coefficient of $y_0$ in $\mf T$. After some computation, we obtain
\al{
\mf T
	&=	-{y_0\ov\sqrt{\paren{\delta\bs V_1}^2+\paren{\delta\bs V_2}^2}}+\cdots,
}
where the dots denote the terms linear in $y_1,\dots, y_5$, which are fixed to be of the order of $\sqrt{\sigma_s}$ by the above Gaussian integrals and are neglected hereafter.
Now the region of the window function $T_\tx{in}<\mf T<T_\tx{out}$ corresponds to
\al{
-\sqrt{\paren{\delta\bs V_1}^2+\paren{\delta\bs V_2}^2}\,T_\tx{out}
	<y_0
	<-\sqrt{\paren{\delta\bs V_1}^2+\paren{\delta\bs V_2}^2}\,T_\tx{in},
}
and the $y_0$ integral yields
\al{
\int\df y_0W\fn{\mf T}
	&=		\sqrt{\paren{\delta\bs V_1}^2+\paren{\delta\bs V_2}^2}
				\paren{T_\tx{out}-T_\tx{in}}.
}

To summarize, the integral over $\paren{\bs X_1,\bs X_2}$ results in\footnote{
When the expression for the probability~\eqref{dP semi final} grows to of order unity as one increases $T_\tx{out}-T_\tx{in}$, one should e.g.\ include the phenomenological factor introduced by Weisskopf and Wigner~\cite{Weisskopf:1930au,Ishikawa:2014ela}.
}
\al{
\df P
	&=	{\kappa^2\ov2}
		{1\ov2E_0}
		{\df^3\bs P_1\ov\paren{2\pi}^{3}2E_1}
		{\df^3\bs P_2\ov\paren{2\pi}^{3}2E_2}
		\paren{2\pi}^4
		\paren{\sqrt{\sigma_t\ov\pi}e^{-\sigma_t\paren{\delta\omega}^2}}
		\paren{\paren{\sigma_s\ov\pi}^{3/2}e^{-\sigma_s\paren{\delta\bs P}^2}}
		\paren{T_\tx{out}-T_\tx{in}}.
		\label{dP semi final}
}
In the wave limit $\sigma_s,\sigma_t\to\infty$, we obtain
\al{
{\df P\ov T_\tx{out}-T_\tx{in}}
	&={\kappa^2\ov2}
		{1\ov2E_0}
		\paren{\df^3\bs P_1\ov\paren{2\pi}^{3}2E_1}
		\paren{\df^3\bs P_2\ov\paren{2\pi}^{3}2E_2}
		\paren{2\pi}^4
		\delta^4\fn{P_1+P_2-P_0}.
			\label{golden rule}
}
This is nothing but the Fermi's golden rule: the decay probability per time-interval $T_\tx{out}-T_\tx{in}$.
The resultant total decay rate reads\footnote{
Let us review the textbook computation: One can use ${\df^3\bs P_a\ov2E_a}=\df^4P_a\,\delta\fn{\paren{P_a}^2+m_\phi^2}\theta\fn{P_a^0}$ and integrate over $\df^4P_2$ to get
\als{
{P\ov T_\tx{out}-T_\tx{in}}
	&=	{\kappa^2\ov4E_0}
		\int{\df^4P_1\ov\paren{2\pi}^{3}}\delta\fn{\paren{P_1}^2+m_\phi^2}\theta\fn{E_1}
		{1\ov\paren{2\pi}^{3}}\delta\fn{\paren{P_0-P_1}^2+m_\phi^2}\theta\fn{E_0-E_1}
		\paren{2\pi}^4\nn
	&=	{\kappa^2\ov4E_0}
		2\pi\int_0^\infty{p_1^2\df p_1\ov\paren{2\pi}^62E_1}\int_{-1}^1\df\cos\theta\,
		\delta\fn{2E_0 E_1-2p_0p_1\cos\theta-m_\Phi^2}\theta\fn{E_0-E_1}
		\paren{2\pi}^4\nn
	&=	{\kappa^2\ov4E_0}
		2\pi\int_{-1\leq{2E_0E_1-m_\Phi^2\ov 2p_0p_1}\leq1}{p_1^2\df p_1\ov\paren{2\pi}^62E_1}
		{1\ov2p_0p_1}\theta\fn{E_0-E_1}
		\paren{2\pi}^4,
}
where $p_1=\ab{\bs P_1}$, $E_1=\sqrt{p_1^2+m_\phi^2}$ and $p_0=\ab{\bs P_0}=\sqrt{E_0^2-m_\Phi^2}$. One may perform the integral in the last line by $p_1\df p_1=E_1\df E_1$ to obtain Eq.~\eqref{total decay rate into scalars}.
}
\al{
{P\ov T_\tx{out}-T_\tx{in}}
	&=	{\kappa^2\ov 32\pi E_0}
		\sqrt{1-{4m_\phi^2\ov m_\Phi^2}}.
			\label{total decay rate into scalars}
}

\subsection{Comments on boundary contribution}\label{boundary comment section}
We examine the contributions~\eqref{G boundary}, which come from either in or out boundary region $\ab{\mf T-T_\tx{in/out}}\lesssim\sigma_t\delta\omega$ (tentatively closing our eyes on the point discussed in footnote~\ref{on boundary effects}). Formulae for the boundary contributions in the boundary limit are summarized in Appendix~\eqref{Boundary limit}.

Let us estimate the effect of the $\df^6y$ integral over the Gaussian peak $e^{-\paren{\mf T-T_\tx{in/out}}^2/\sigma_t}$ in Eq~\eqref{G limit}, which results from the limit ${\paren{\mf T-T_\tx{in/out}}^2\ov\sigma_t}+\sigma_t\paren{\delta\omega}^2\gg1$:
\al{
\df P	
	&\to
		{\kappa^2\ov2}
		{1\ov2E_0}
		{\df^3\bs P_1\ov\paren{2\pi}^{3}2E_1}
		{\df^3\bs P_2\ov\paren{2\pi}^{3}2E_2}
		\paren{2\pi}^4
		\paren{\sqrt{\sigma_t\ov\pi}}
		\paren{\paren{\sigma_s\ov\pi}^{3/2}e^{-\sigma_s\paren{\delta\bs P}^2}}
			\nn
	&\qquad\times
		\sqrt{
		{\sigma_t\ov\pi^5}
		\paren{\sigma_s\ov\sigma_0\sigma_1\sigma_2}^3}\,
		\df^6y\,
		e^{-\mc R}
		e^{-{\paren{\mf T-T_\tx{in/out}}^2\ov\sigma_t}}
		{2\ov\pi}
		{1\ov{\paren{\mf T-T_\tx{in/out}}^2\ov\sigma_t}+\sigma_t\paren{\delta\omega}^2}.
		\label{dP limit taken}
}
As discussed in the paragraph containing Eq.~\eqref{opposite limit}, this expression is valid only when $\sigma_t\paren{\delta\omega}^2\gg1$ at $\mf T=T_\tx{in/out}$; see Appendix~\ref{Boundary limit} for possible generalization.

\emph{Naively,} the integral over the above-mentioned Gaussian peak would be estimated by taking the formal limit $\sigma_t\to0$,\footnote{
It should be understood that the limit $\sigma_t\to0$ is taken with fixed $\sqrt{\sigma_t}\delta\omega$.
}
\al{
e^{-{\paren{\mf T-T_\tx{in/out}}^2\ov\sigma_t}}
	&\to	\sqrt{\pi\sigma_t}\,\delta\fn{\mf T-T_\tx{in/out}},
		\label{delta function limit}
}
and by regarding the integral $\df^5y\,e^{-\mc R}$ as Gaussian~\eqref{five dimensional integral}; the remaining $y_0$ integral would again give the factor $\sqrt{\paren{\delta\bs V_1}^2+\paren{\delta\bs V_2}^2}$:
\al{
\df P
	&\sim	{\kappa^2\ov2}
		{1\ov2E_0}
		{\df^3\bs P_1\ov\paren{2\pi}^{3}2E_1}
		{\df^3\bs P_2\ov\paren{2\pi}^{3}2E_2}
		\paren{2\pi}^4
		\sqbr{\paren{\sigma_s\ov\pi}^{3/2}e^{-\sigma_s\paren{\delta\bs P}^2}}
		{2\ov\pi}
		{1\ov\paren{\delta\omega}^2}.
}
We may further take the plane-wave limit $\sigma_s\to\infty$, which renders the factor in the square brackets into the delta function $\delta^3\fn{\delta\bs P}$:
\al{
\df P
	&=	{\kappa^2\ov2}
		{1\ov2E_0}
		{\df^3\bs P_1\ov\paren{2\pi}^{3}2E_1}
		{1\ov\paren{2\pi}^{3}2E_2}
		\paren{2\pi}^4
		{2\ov\pi}
		{1\ov\paren{\delta E}^2},
}
where we have also replaced $\delta\omega$ by $\delta E$; see Eq.~\eqref{energy shift etc}.
We see that the ultraviolet behavior of the momentum integral is
\al{
\df P
	&\propto \int{\df\ab{\bs P_1}\ov\ab{\bs P_1}^2},	
			\label{no divergence}
}
which is convergent. This convergence itself is independent of the limits that we have taken.

There is no ultraviolet divergence from the boundary regions if the decay is due to the superrenormalizable interaction~\eqref{superrenormalizable interaction}. In contrast, if the decay of scalar were due to a marginal operator of dimension four, we would have got a linearly divergent integral instead of Eq~\eqref{no divergence}.\footnote{
Naively, the dimensional analysis tells that the tree-level \red{two-body} decay of a scalar due to a dimension-$d$ operator would result in the ultraviolet divergence of the order of $2d-7$. This is the case for the non-renormalizable interaction~\eqref{diphoton interaction} too.
}

We comment on the possible ultraviolet divergence at the boundary. 
First, one might want to take into account the ``uncertainty'' of $T_\tx{in/out}$ that is defined in our treatment to be the time \red{(}at which the interacting state can well be identified to the free state\red{)}, by ``diffusing the boundary'' \`a la Stueckelberg~\cite{Stueckelberg:1951zz}. This would provide an additional UV suppression factor on the momentum integral, but the necessary unitarity violation requires the change of very definition of the $S$-matrix.
Second, as said in footnote~\ref{on boundary effects}, the identification of the interacting state with the free state at $T_\tx{in/out}$ cannot be justified for the boundary contribution.
Third, in realistic (particle physics) situation, there is no ideally-sharp time boundary but some production and detection mechanisms that are extended in spacetime. The phenomenology on the boundary region could strongly depend on the microscopic physics of the boundary. 
Thus, the boundary contribution depends on the situation or \red{might not be valid} 
when it is ultraviolet divergent. 
Further discussion and implication will be presented elsewhere.

\section{Diphoton decay}\label{diphoton section}
In order to exhibit how to generalize the simplest scalar decay by the interaction~\eqref{superrenormalizable interaction} to more realistic cases, we consider the decay of a pseudoscalar into a diphoton pair:
\al{
\mc L_\tx{int}
	&=	-{g_\Phi\ov4}\Phi\varepsilon_{\lambda\mu\nu\rho}F^{\lambda\mu}F^{\nu\rho}
	=	-\mc H_\tx{int},
		\label{diphoton interaction}
}
where $\varepsilon^{\lambda\mu\nu\rho}$ is the totally anti-symmetric tensor and $g_\Phi$ is a coupling constant of mass dimension~$-1$. For the pion decay, we set
$g_\Phi 
={\sqrt{2}\alpha\ov\pi f_\pi}$, 
where $\alpha\simeq 1/137$ and $f_\pi\simeq 130\,\tx{MeV}$ are the fine-structure and pion decay constants, respectively.

It is actually straightforward to generalize the previous analysis to the diphoton decay.
The photon field operator can be expanded in terms of the creation/annihilation operators of the plane and Gaussian waves as
\al{
\h A_\mu\fn{x}
	&=	\left.\sum_s\int{\df^3\bs p\ov\sqrt{2p^0}\paren{2\pi}^{3/2}}
			\sqbr{
				\h a\fn{s,\bs p}\epsilon_\mu\fn{s,\bs p}e^{ip\cdot x}
				+\h a^\dagger\fn{s,\bs p}\epsilon^*_\mu\fn{s,\bs p}e^{-ip\cdot x}
				}\right|_{p^0=\ab{\bs p}}\\
	&=	\sum_s\int{\df^3\bs X\df^3\bs P\ov\paren{2\pi}^3}
			\sqbr{
				E_\mu\fn{s,\sigma;X,\bs P}\h A\fn{s,\sigma;X,\bs P}
				+E_\mu^*\fn{s,\sigma;X,\bs P}\h A^\dagger\fn{s,\sigma;X,\bs P}
				},
}
respectively, where
\al{
E_\mu\fn{s,\sigma;X,\bs P}
	&=	\left.\int{\df^3\bs p\ov\sqrt{2p^0}}\epsilon_\mu\fn{s,\bs p}e^{ip\cdot\paren{x-X}-{\sigma\ov2}\paren{\bs p-\bs P}^2}\right|_{p^0=\ab{\bs p}};
}
see Appendix~\ref{Gaussian wave packets section}. The saddle-point approximation in the large-width expansion gives\footnote{
One can explicitly check that the next-leading order terms in the expansion~\eqref{Ps is P plus delta} cancel out in the final expression of the probability $\df P$. For example, the saddle-point momentum remains massless at the next-leading order: $\paren{\bs P\pm i{\bs x-\bs \Xi\fn{t}\ov\sigma}}^2=\bs P^2+\Or{1\ov\sigma^2}$.
}
\al{
E_\mu\fn{s,\sigma;X,\bs P}
	&\simeq
		\paren{\sigma\ov\pi}^{4/3}\paren{2\pi\ov\sigma}^{3/2}\paren{1\ov\paren{2\pi}^32\ab{\bs P}}^{1/2}
		\epsilon_\mu\fn{s,\bs P}
		e^{-i\ab{\bs P}\paren{t-T}+i\bs P\cdot\paren{\bs x-\bs X}-{\paren{\bs x-\bs \Xi\fn{t}}^2\ov2\sigma}};
			\label{vector saddle point}
}
see Appendix~\ref{saddle point section}. 

In obtaining the $S$-matrix, all we have to do is to replace $\kappa$ by
\al{
\kappa
	&\mapsto	g_\Phi\,\varepsilon_{\lambda\mu\nu\rho}
			P_1^\lambda
			\epsilon^{\mu*}\fn{s_1,\bs P_1}
			P_2^\nu
			\epsilon^{\rho*}\fn{s_2,\bs P_2}
}
in Eq.~\eqref{S Gauss-integrated}.
The spin-summed decay probability is then, from Eq.~\eqref{dP full},
\al{
\df P
	&=	
		{2g_\Phi^2\paren{P_1\cdot P_2}^2\ov2}
		{1\ov2E_0}
		{\df^3\bs P_1\ov\paren{2\pi}^{3}2E_1}
		{\df^3\bs P_2\ov\paren{2\pi}^{3}2E_2}
		\paren{2\pi}^4
		\paren{\sqrt{\sigma_t\ov\pi}e^{-\sigma_t\paren{\delta\omega}^2}}
		\paren{\paren{\sigma_s\ov\pi}^{3/2}e^{-\sigma_s\paren{\delta\bs P}^2}}
			\nn
	&\qquad\times
		\sqrt{
		{\sigma_t\ov\pi^5}
		\paren{\sigma_s\ov\sigma_0\sigma_1\sigma_2}^3}\,
		\df^3\bs X_1\,\df^3\bs X_2\,
		e^{-\mc R}\,\ab{G\fn{\mf T}}^2.
}
where we have used
\al{
\sum_{s_1,s_2}\ab{\varepsilon_{\lambda\mu\nu\rho}
			P_1^\lambda
			\epsilon^{\mu}\fn{s_1,\bs P_1}
			P_2^\nu
			\epsilon^{\rho}\fn{s_2,\bs P_2}
			}^2
			&=	2\paren{P_1\cdot P_2}^2.
}

After taking the plane-wave limit, the final expression for the Fermi's golden rule~\eqref{golden rule} becomes
\al{
{\df P\ov T_\tx{out}-T_\tx{in}}
	&=
		{g_\Phi^2m_\Phi^4\ov 8E_0}
		\paren{\df^3\bs P_1\ov\paren{2\pi}^{3}2E_1}
		\paren{\df^3\bs P_2\ov\paren{2\pi}^{3}2E_2}
		\paren{2\pi}^4
		\delta^4\fn{P_1+P_2-P_0},
}
where we used, under the momentum delta function and the on-shell condition,
\al{
P_1\cdot P_2
	&=	{\paren{P_1+P_2}^2\ov2}
	=	{P_0^2\ov2}
	=	-{m_\Phi^2\ov2}.
}
The total decay rate is
\al{
{P\ov T_\tx{out}-T_\tx{in}}
	&=	{g_\Phi^2m_\Phi^4\ov 64\pi E_0}.
}
That is, the replacement in the final expression reads $\kappa^2\mapsto g_\Phi^2m_\Phi^4/2$ (and of course $m_\phi\mapsto0$).

\section{Summary}\label{summary}
We have reformulated the Gaussian $S$-matrix \red{within a finite time interval} in the Gaussian wave-packet formalism.
The normalizable Gaussian basis allows the computation of the decay probability without the \red{momentum-space $\delta^4\fn{0}$} singularity that necessarily appears in the one involving the plane-wave basis. 
We have performed the exact four dimensional integration over the interaction point~$x$ for the decay probability\red{. 
The unitarity is manifestly maintained throughout the whole computation.}

We have proposed a separation of the obtained result into the bulk and boundary parts.
This separation corresponds to whether the interaction point is near the time boundary or not and hence is rather intuitive and easy to envisage.
The Fermi's golden rule is derived from the bulk contribution.
\red{As a byproduct, we have also shown that} the ultraviolet divergence in the boundary contribution is absent for the decay of a scalar into a pair of light scalars by the superrenormalizable interaction\red{, though its physical significance is yet to be confirmed}.
We have generalized our results to the case of diphoton decay and to more general initial and final state particles.

\subsection*{Acknowledgement}
We owe Hiromasa Nakatsuka for valuable contributions at the early stages of this work and for reading the manuscript.
We are grateful to Osamu Jinnouchi, Arisa Kubota, Terence Sloan, and Risa Ushioda for stimulating discussion. We thank Akio Hosoya, Izumi Ojima, and Masaharu Tanabashi for useful comments.
The work of K.I.\ and K.O.\ are in part supported by JSPS KAKENHI Grant Nos.~24340043 (K.I.) and 15K05053 (K.O.).
\appendix
\section*{Appendix}
\section{Gaussian wave packet formalism}\label{GWP section}
We review and spell out our notation for the Gaussian wave-packet basis~\cite{Ishikawa:2005zc,Ishikawa:2014eoa,Ishikawa:2014uma}.
\subsection{Heisenberg, Schr\"odinger, and interaction pictures}\label{pictures}
We may always separate the total Lagrangian density $\mc L$ into the free part $\mc L_\tx{free}$ that contains quadratic terms in fields and the interaction one $\mc L_\tx{int}$ that is the rest:
\al{
\mc L
	&=	\mc L_\tx{free}+\mc L_\tx{int}.
}
Correspondingly, we may separate the Hamiltonian (density) $H$ ($\mc H$) into the free and interaction parts $H_\tx{free}$ ($\mc H_\tx{free}$) and $H_\tx{int}$ ($\mc H_\tx{int}$), respectively:
\al{
H
	&=	H_\tx{free}+H_\tx{int},	&
\mc H
	&=	\mc H_\tx{free}+\mc H_\tx{int}.
}
We list the time dependence of the physical state, operator, and eigenbasis in the Heisenberg, Schr\"odinger, and interaction pictures in the following table:\footnote{
We choose our reference time to identify these three pictures to be $t=0$ throughout this paper.
}
\bigskip\\
\begin{tabular}{|c||rlrlrl|}
\hline
\tx{Picture}
&
&\hspace{-11pt}\tx{State}&
&\hspace{-11pt}\tx{Operator} &
&\hspace{-11pt}\tx{Basis}\\
\hline
\tx{Heisenberg}&
&\hspace{-11pt}$\Ket{\Phi}\Hpr$&
$\h O^\Hp\fn{t}$
	&\hspace{-11pt}$=	e^{i\h Ht}\h O^\Sp e^{-i\h Ht}$ &
$\Ket{\Phi,t}\HBr$
	&\hspace{-11pt}$=	e^{i\h Ht}\Ket{\Phi}\SBr$\\
\tx{Schr\"odinger}& 
$\Ket{\Phi,t}\Spr$
	&\hspace{-11pt}$=	e^{-i\h Ht}\Ket{\Phi}\Hpr$ & 
&\hspace{-11pt}$\h O^\Sp$&
&\hspace{-11pt}$\Ket{\Phi}\SBr$
		\\
\tx{Interaction} & 
$\Ket{\Phi,t}\Ipr$
	&\hspace{-11pt}$=	e^{i\h H_\tx{free}t}e^{-i\h Ht}\Ket{\Phi}\Hpr$ & 
$\h O^\Ip\fn{t}$
	&\hspace{-11pt}$=	e^{i\h H_\tx{free}t}\h O^\Sp e^{-i\h H_\tx{free}t}$ &
$\Ket{\Phi,t}\IBr$
	&\hspace{-11pt}$=	e^{i\h H_\tx{free}t}\Ket{\Phi}\SBr$\\
\hline
\end{tabular}
\smallskip\\
Throughout this paper, $\h H$ ($\h H_\tx{free}$) denotes the time-independent total (free) Hamiltonian in the Schr\"odinger or Heisenberg (interaction) picture.
Any Schr\"odinger eigenbasis $\Ket{\Phi}\SBr$ can be regarded as a Heisenberg state:
\al{
\Ket{\Phi}\SBr
	&=	\Ket{\Phi}\Hpr.
}

\subsection{Plane-wave expansion}\label{plane wave expansion section}
Let us spell out the ordinary plane-wave basis as a preparation for the Gaussian basis.

A free field operator $\h\Psi^\Ip\fn{x}$ at $x=\paren{x^0,\bs x}=\paren{t,\bs x}$ in the interaction picture can be expanded in terms of the plane waves $e^{\pm ip\cdot x}$:
\al{
\h\Psi^\Ip\fn{x}
	&=	\left.
			\sum_s
			\int
			{\df^3\bs p\ov\sqrt{2p^0}\paren{2\pi}^{3/2}}
			\sqbr{
				\h a_\Psi\fn{s,\bs p}U\fn{s,\bs p}e^{ip\cdot x}
				+\h a_\Psi^{c\dagger}\fn{s,\bs p}V\fn{s,\bs p}e^{-ip\cdot x}
				}
		\,
		\right|_{p^0=E_\Psi\fn{\bs p}},
}
where $E_\Psi\fn{\bs p}$ is given in Eq.~\eqref{on-shell energy};
$s$ is the helicity or the spin (of the little group);
and the coefficient functions $U$ and $V$ are given, e.g., for a scalar ($s=0$), a Dirac spinor ($s=\pm1/2$), and a massless vector ($s=\pm1$) as\footnote{
The dependence of $u$ and $v$ on the mass $m_\Psi$ is made implicit.
}
\al{
U\fn{s,\bs p}
	&=	\begin{cases}
		1,\\
		u\fn{s,\bs p},\\
		\epsilon_\mu\fn{s,\bs p},
		\end{cases}
		&
V\fn{s,\bs p}
	&=	\begin{cases}
		1,\\
		v\fn{s,\bs p},
			\\
		\epsilon_\mu^*\fn{s,\bs p},
		\end{cases}
		&
&\quad\tx{for}\quad&
\h\Psi^\Ip\fn{x}
	&=	\begin{cases}
		\h\varphi\fn{x}	&	\tx{(scalar)},\\
		\h\psi\fn{x}&	\tx{(spinor)},\\
		\h A_\mu\fn{x}&	\tx{(vector)}.
		\end{cases}
}
Here and hereafter, the annihilation operators $\h a$ and $\h a^c$ are always given in the Schr\"odinger picture (i.e.\ time-independently) as usual.
The creation and annihilation operators obey
\al{
\commutator{\h a_\Psi\fn{s,\bs p}}{\h a_{\Psi'}^\dagger\fn{s',\bs p'}}_\pm
	&=	\delta_{\Psi\Psi'}\delta_{ss'}\delta^3\fn{\bs p-\bs p'},	\nn
\commutator{\h a_\Psi^c\fn{s,\bs p}}{\h a_{\Psi'}^{c\dagger}\fn{s',\bs p'}}_\pm
	&=	\delta_{\Psi\Psi'}\delta_{ss'}\delta^3\fn{\bs p-\bs p'},\nn
\tx{others}
	&=	0.
}
where plus and minus signs correspond to the anticommutator and commutator when both $\Psi$ and $\Psi'$ are fermions ($s=\pm1/2$) and when otherwise, respectively.
A real (Majorana) field corresponds to $\h a^c\fn{s,\bs p}=\h a\fn{s,\bs p}$.

A free massless (massive) one-particle state with a definite helicity (spin) $s$ and a momentum $\bs p$ is given by
\al{
\Ket{s,\bs p}\SBrP
	&=	\h a_\Psi^\dagger\fn{s,\bs p}\Ket{0},
		\label{s p state}
}
where we have normalized such that
\al{
\SBlP\Braket{s,\bs p|s',\bs p'}\SBrPd
	&=	\delta^3\fn{\bs p-\bs p'}\delta_{ss'}\delta_{\Psi\Psi'},	&
\int\df^3\bs p\,\Ket{s,\bs p}\SBrP\SBlP\Bra{s,\bs p}
	&=	\h 1,
}
where $\h 1$ is the identity operator in the one-particle subspace with a definite $s$.
One obtains the free Hamiltonian
\al{
\h H_\tx{free}=\sum_\Psi\sum_s\int\df^3\bs p\,E_\Psi\fn{\bs p}\h a_\Psi^\dagger\fn{s,\bs p}\h a_\Psi\fn{s,\bs p}
}
up to a constant term, and the state~\eqref{s p state} becomes the eigenbasis for it:
\al{
\h H_\tx{free}\Ket{s,\bs p}\SBrP
	&=	E_\Psi\fn{\bs p}\Ket{s,\bs p}\SBrP.
}

As in the ordinary quantum mechanics, the one-particle position eigenbasis $\!\SBlP\Bra{s,\bs x}$ is defined to yield the plane-wave function when multiplied on $\Ket{s,\bs p}\SBrP$:
\al{
\SBlP\Braket{s,\bs x|s',\bs p}\SBrPd
	&=	{e^{i\bs p\cdot\bs x}\ov\paren{2\pi}^{3/2}}\delta_{ss'}\delta_{\Psi\Psi'},
}
where its normalization is chosen such that
\al{
\SBlP\Braket{s,\bs x|s',\bs x'}\SBrPd
	&=	\delta^3\fn{\bs x-\bs x'}\delta_{ss'}\delta_{\Psi\Psi'},	&
\int\df^3\bs x\,\Ket{s,\bs x}\SBrP\SBlP\Bra{s,\bs x}
	&=	\h 1.
}
We may call the position eigenbasis in the interaction picture at time $t$ ``the time-translated position eigenbasis at $x=\paren{x^0,\bs x}=\paren{t,\bs x}$'':
\al{
\IBlP\Bra{s,x}
	&:=	\SBlP\Bra{s,\bs x}e^{-i\h H_\tx{free}t}.
		\label{x-state defined}
}
Concretely, we get
\al{
\IBlP\Braket{s,x|s',\bs p}\SBrP
	&=	
		\left.{e^{ip\cdot x}\ov\paren{2\pi}^{3/2}}\right|_{p^0=E_\Psi\fn{\bs p}}
		\delta_{ss'}.
		\label{concrete x-state}
}
The completeness still holds,
\al{
\int\df^3\bs x\Ket{s,x}\IBrP\IBlP\Bra{s,x}
	&=	\h 1,
}
whereas the orthogonality holds only at the equal time:
\al{
\left.\IBlP\Braket{s,x|s',x'}\IBrPd\right|_{t=t'}
	&=	\delta^3\fn{\bs x-\bs x'}\delta_{ss'}\delta_{\Psi\Psi'}.
}
Now we may rewrite
\al{
\h\Psi^\Ip\fn{x}
	&=	\sum_s
		\int{\df^3\bs p\ov\sqrt{2E_\Psi\fn{\bs p}}}
		\sqbr{
			\h a\fn{s,\bs p}U\fn{s,\bs p}\paren{\IBlP\Braket{s,x|s,\bs p}\SBrP}
			+\h a^{c\dagger}\fn{s,\bs p}V\fn{s,\bs p}
				\paren{\IBlP\Braket{s,x|s,\bs p}\SBrP}^*
			}.
		\label{expansion by plane wave}
}

\subsection{Gaussian wave packets}\label{Gaussian wave packets section}

We define a free Gaussian wave-packet state $\Ket{s,\sigma;\bs X,\bs P}\SBrP$ that is localized at $\bs X$ with the width $\sqrt{\sigma}$ and with the central momentum $\bs P$ by the standard Gaussian wave function of $\bs x$:
\al{
\SBlP\Braket{s,\bs x|s',\sigma;\bs X,\bs P}\SBrP
	&=	{1\ov\paren{\pi\sigma}^{3/4}}e^{i\bs P\cdot\paren{\bs x-\bs X}}e^{-{1\ov2\sigma}\paren{\bs x-\bs X}^2}\delta_{ss'},
		\label{Gaussian wave function at zero}
}
where we have normalized such that
\al{
\int\df^3\bs x\ab{\SBlP\Braket{s,\bs x|s,\sigma;\bs X,\bs P}\SBrP}^2=1.
}
Analogously to the plane-wave basis in Eq.~\eqref{x-state defined}, we may define the Gaussian basis that is centered at $X=\paren{X^0,\bs X}=\paren{T,\bs X}$ by
\al{
\IBlP\Bra{s,\sigma;X,\bs P}
	&:=	\SBlP\Bra{s,\sigma;\bs X,\bs P}e^{-i\h H_\tx{free}T}.
}
Concretely, we obtain
\al{
\IBlP\Braket{s,\sigma;X,\bs P|s,\bs p}\SBr
	&=	\paren{\sigma\ov\pi}^{3/4}
		\left.
		e^{ip\cdot X}e^{-{\sigma\ov2}\paren{\bs p-\bs P}^2}
		\right|_{p^0=E_\Psi\fn{\bs p}},
		\label{concrete Gaussian wave function}
}
where we have used Eqs.~\eqref{concrete x-state} and \eqref{Gaussian wave function at zero}.\footnote{Though not quite useful, we may also write down the time-shifted Gaussian wave function in an integral form:
\als{
\IBlP\Braket{s,x|s,\sigma;X,\bs P}\IBrP
	&=	\paren{\sigma\ov\pi}^{3/4}
		\int{\df^3\bs p\ov\paren{2\pi}^{3/2}}
		\left.
		e^{ip\cdot\paren{x-X}-{\sigma\ov2}\paren{\bs p-\bs P}^2}
		\right|_{p^0=E_\Psi\fn{\bs p}}.
}
}
Note that the completeness relation now becomes\footnote{
One can explicitly show that
\als{
\Bra{\bs p}\paren{\int{\df^3\bs X\,\df^3\bs P\ov\paren{2\pi}^3}\Ket{\sigma;X,\bs P}\Bra{\sigma;X,\bs P}}\Ket{\bs p'}
	=	\delta^3\fn{\bs p-\bs p'},
}
where we have tentatively omitted $\Psi$, $s$, etc.
}
\al{
\int{\df^3\bs X\,\df^3\bs P\ov\paren{2\pi}^3}\Ket{s,\sigma;X,\bs P}\IBrP\IBlP\Bra{s,\sigma;X,\bs P}
	&=	\h 1
		\label{wave packet completeness}
}
and that the Gaussian basis states are not orthogonal to each other even if $T=T'$:
\al{
\lefteqn{\left.\IBlP\Braket{s,\sigma;X,\bs P|s',\sigma';X',\bs P'}\IBrPd\right|_{T=T'}}\nn
	&\qquad=	
		\paren{\sigma_\tx{I}\ov\sigma_\tx{A}}^{3/4}
		e^{-{1\ov4\sigma_\tx{A}}\paren{\bs X-\bs X'}^2}
		e^{-{\sigma_\tx{I}\ov4}\paren{\bs P-\bs P'}^2}
		e^{{i\ov2\sigma_\tx{I}}\paren{\sigma\bs P+\sigma'\bs P'}\cdot\paren{\bs X-\bs X'}}\,
		\delta_{ss'}\delta_{\Psi\Psi'},
					\label{non-orthogonality}
}
where $\sigma_\tx{A}:={\sigma+\sigma'\ov2}$ and $\sigma_\tx{I}:=\paren{\sigma^{-1}+\sigma^{\pr-1}\ov2}^{-1}={2\sigma\sigma'\ov\sigma+\sigma'}$ are the average and the inverse of inverse average, respectively.
Namely, the Gaussian basis is overcomplete.

Now we define the creation operator of the free wave packet $\h A_\Psi^\dagger\fn{s,\sigma;X,\bs P}$ by\footnote{
We note that, in the Gaussian formulation, the postulation~(c) in Ref.~\cite{Newton:1949cq} does not hold, nor its conclusion of no-go, because the Gaussian basis states are not orthogonal to each other even when their location $\bs X$ and $\bs X'$ are different, as can be seen in Eq.~\eqref{non-orthogonality}. We thank Akio Hosoya and Izumi Ojima for pointing out this issue.
}
\al{
\h A_\Psi^\dagger\fn{s,\sigma;X,\bs P}\Ket{0}
	&=	\Ket{s,\sigma;X,\bs P}\IBrP,
}
which leads to\footnote{
When we expand $\h A$ by $\h a$ as $\h A\fn{\sigma;X,\bs P}=\int\df^3\bs p'\,f_{\bs p'}\fn{\sigma;X,\bs P}\,\h a\fn{\bs p'}$ (we have omitted $\Psi$, $s$, etc.), we get
\als{
\Bra{0}\h A\fn{\sigma;X,\bs P}\Ket{\bs p}
	&=	\Bra{0}\int\df^3\bs p'\,f_{\bs p'}\fn{\sigma;X,\bs P}\,\h a\fn{\bs p'}\Ket{\bs p}
	=	f_{\bs p}\fn{\sigma;X,\bs P},
}
which is equated to Eq.~\eqref{concrete Gaussian wave function} to yield Eq.~\eqref{A in terms of a}.
}
\al{
\h A_\Psi\fn{s,\sigma;X,\bs P}
	&=	\int\df^3\bs p
		\paren{\IBlP\Braket{s,\sigma;X,\bs P|s;\bs p}\SBrP}
		\h a_\Psi\fn{s,\bs p},
		\label{A in terms of a}\\
\h a_\Psi\fn{s,\bs p}
	&=	\int{\df^3\bs X\df^3\bs P\ov\paren{2\pi}^3}
		\paren{\SBlP\Braket{s,\bs p|s,\sigma;X,\bs P}\IBrP}
		\h A_\Psi\fn{s,\sigma;X,\bs P}.
		\label{a in terms of A}
}
Note that
\al{
\sqbr{\h A_\Psi\fn{s,\sigma;X,\bs P},\h A_{\Psi'}^\dagger\fn{s',\sigma';X',\bs P'}}_\pm
	&=	\IBlP\Braket{s',\sigma;X,\bs P|s,\sigma';X',\bs P'}\IBrPd,\nn
\tx{others}
	&=	0.
}

To obtain the explicit form of the expansion in terms of the Gaussian basis, one may put Eq.~\eqref{a in terms of A} into Eq.~\eqref{expansion by plane wave}:
\al{
\h\Psi^\Ip\fn{x}
	&=	\sum_s\int{\df^3\bs X\df^3\bs P\ov\paren{2\pi}^3}
		\sqbr{
			\mc U_{s,\sigma;X,\bs P}\fn{x}\h A_\Psi\fn{s,\sigma;X,\bs P}
			+\mc V_{s,\sigma;X,\bs P}\fn{x}\h A_\Psi^{c\dagger}\fn{s,\sigma;X,\bs P}
			},
			\label{Gaussian expansion}
}
where
\al{
\mc U_{s,\sigma;X,\bs P}\fn{x}
	&:=	\int{\df^3\bs p\ov\sqrt{2E_\Psi\fn{\bs p}}}
		\paren{\IBlP\Braket{s,x|s,\bs p}\SBrP}
		U\fn{s,\bs p}
		\paren{\SBlP\Braket{s,\bs p|s,\sigma;X,\bs P}\IBrP},
			\label{curly U defined}\\
\mc V_{s,\sigma;X,\bs P}\fn{x}
	&:=	\int{\df^3\bs p\ov\sqrt{2E_\Psi\fn{\bs p}}}
		\paren{\IBlP\Braket{s,\sigma;X,\bs P|s,\bs p}\SBrP}
		V\fn{s,\bs p}
		\paren{\SBlP\Braket{s,\bs p|s,x}\IBrP}.
			\label{curly V defined}
}
Using Eqs.~\eqref{concrete x-state} and \eqref{concrete Gaussian wave function}, one may write down the integral form more explicitly:
\al{
\mc U_{s,\sigma;X,\bs P}\fn{x}
	&=	
		\left.
		\paren{\sigma\ov\pi}^{3/4}\int{\df^3\bs p\ov\sqrt{2p^0}\paren{2\pi}^{3/2}}
		U\fn{s,\bs p}e^{ip\cdot\paren{x-X}-{\sigma\ov2}\paren{\bs p-\bs P}^2}
		\right|_{p^0=E_\Psi\fn{\bs p}},
			\label{curly U}\\
\mc V_{s,\sigma;X,\bs P}\fn{x}
	&=	
		\left.
		\paren{\sigma\ov\pi}^{3/4}\int{\df^3\bs p\ov\sqrt{2p^0}\paren{2\pi}^{3/2}}
		V\fn{s,\bs p}e^{-ip\cdot\paren{x-X}-{\sigma\ov2}\paren{\bs p-\bs P}^2}
		\right|_{p^0=E_\Psi\fn{\bs p}}.
		\label{curly V}
}
Note that $T$ ($=X^0$) and $\sigma$ can be chosen arbitrarily for the expansion~\eqref{Gaussian expansion}.
The coefficient functions $\mc U$ and $\mc V$ are nothing but the external line factor in the computation of $S$-matrix:
\al{
\Bra{0}\h\Psi^\Ip\fn{x}\Ket{s,\sigma;X,\bs P}\IBrP
	&=	\sum_s\int\df^6\bs\Pi'\ 
			\mc U_{s;\Pi'}\fn{x}\,
			\Bra{0}\h A_\Psi\fn{s;\Pi'}\h A^\dagger_\Psi\fn{s;\Pi}\Ket{0}
			\nn
	&=	\sum_s\int\df^6\bs\Pi'
			\paren{\int{\df^3\bs p\ov\sqrt{2E_\Psi\fn{\bs p}}}
			\Braket{x|\bs p}U\fn{\bs p}\Braket{\bs p|\Pi'}}
			\Braket{\Pi'|\Pi}\nn
	&=	\sum_s
			\int{\df^3\bs p\ov\sqrt{2E_\Psi\fn{\bs p}}}
			\Braket{x|\bs p}U\fn{\bs p}\Braket{\bs p|\Pi}\nn
	&=	\mc U_{s,\sigma;X,\bs P}\fn{x},
}
and so on, where we have omitted $\Psi$, $\sigma$, and $s$ in the intermediate steps and have used the abbreviation~\eqref{shorthand notation}.

\section{Saddle-point approximation}\label{saddle point section}
Let us obtain the approximate formulae for the functions~\eqref{curly U} and \eqref{curly V} using the saddle-point method for the large width expansion. When evaluating the momentum integration, we encounter the exponent of the form\footnote{
In taking the large $\sigma$ expansion, we have to be careful about the region of large $\ab{\bs x-\bs X}$ and/or large $\ab{t-T}$. Here we assume that we are in a generic non-singular point in the parameter space in which the contribution from such an interaction point $x$ is suppressed and that the large $\sigma$ expansion works.
}
\al{
F_\pm\fn{\bs p}
	&:=	\mp iE\fn{\bs p}\paren{t-T}\pm i\bs p\cdot\paren{\bs x-\bs X}-{\sigma\ov2}\paren{\bs p-\bs P}^2,
}
where $E\fn{\bs p}:=\sqrt{\bs p^2+m^2}$.
First,
\al{
{\p F_\pm\fn{\bs p}\over\p p_i}
	&=	\mp iv_i\fn{\bs p}\paren{t-T}
			\pm i\paren{x-X}_i
			-\sigma\paren{p-P}_i,	\label{first derivative}\\
{\p^2 F_\pm\fn{\bs p}\over\p p_i\p p_j}
	&=	\mp i{t-T\over E\fn{\bs p}}\sqbr{\delta_{ij}-v_i\fn{\bs p}v_j\fn{\bs p}}
			-\sigma\delta_{ij},
}
where
\al{
\bs v\fn{\bs p}
	&:=	{\bs p\ov E\fn{\bs p}}
}
and we have used
\al{
{\p E\fn{\bs p}\ov\p p_j}
	&=	v_j\fn{\bs p},	&
{\p v_i\fn{\bs p}\over\p p_j}
	&=	{\delta_{ij}-v_i\fn{\bs p}v_j\fn{\bs p}\over E\fn{\bs p}}.
}

Let $\bs P_s$ be the solution to the saddle point condition:
\al{
{\p F_\pm\fn{\bs P_s}\ov\p p_i}
	&=	0,	\label{saddle point condition}
}
where for arbitrary $\bs P$ and function $f\fn{\bs p}$, we write
\al{
{\p f\fn{\bs P}\ov\p p_i}
	&:=	\left.{\p f\fn{\bs p}\ov\p p_i}\right|_{\bs p=\bs P}.
}
The zeroth and second derivatives read
\al{
F_\pm\fn{\bs P_s}
	&=	\mp iE\fn{\bs P_s}\paren{t-T}\pm i\bs P_s\cdot\paren{\bs x-\bs X}-{\sigma\ov2}\paren{\bs P_s-\bs P}^2,
		\\
{\p^2F_\pm\fn{\bs P_s}\ov\p p_i\p p_j}
	&=	-\paren{\sigma\pm i{t-T\ov E\fn{\bs P_s}}}\delta_{ij}
		\pm i{t-T\ov E\fn{\bs P_s}}v_i\fn{\bs P_s}v_j\fn{\bs P_s}
	=:	M_{ij}.
}
The complex symmetric matrix $M_{ij}=a\delta_{ij}+bv_jv_j$ can be diagonalized by a complex special orthogonal matrix $U$ that obeys $U^\t U=1$ and $\det U=1$:\footnote{
Explicitly, one may e.g.\ take
\als{
U
	&=	\bmat{
			v_1v_3\over\sqrt{v_1^2+v_2^2}\sqrt{\bs v^2}&
				-{v_2\over\sqrt{v_1^2+v_2^2}}&
				v_1\over\sqrt{\bs v^2}\\
			v_2v_3\over\sqrt{v_1^2+v_2^2}\sqrt{\bs v^2}&
				{v_1\over\sqrt{v_1^2+v_2^2}}&
					v_2\over\sqrt{\bs v^2}\\
			-{\sqrt{v_1^2+v_2^2}\over\sqrt{\bs v^2}}&
				0&
					v_3\over\sqrt{\bs v^2}
			}.
}
}
\al{
U^\t MU
	&=	\bmat{a&0&0\\ 0&a&0\\ 0&0&a+b\bs v^2}.
}
The complex Gaussian integral reads
\al{
\int\df^3p\,e^{F_\pm\fn{\bs p}}Q\fn{\bs p}
	&\simeq
		e^{F\fn{\bs P_s}}Q\fn{\bs P_s}\int\df^3p\,e^{{1\ov2}\paren{p-P_s}_i{\p^2F\fn{\bs P_s}\ov\p p_i\p p_j}\paren{p-P_s}_j}\nn
	&=	\paren{2\pi\ov\sigma}^{3/2}
		{e^{F\fn{\bs P_s}}Q\fn{\bs P_s}\ov
			\paren{1\pm i{t-T\ov \sigma E\fn{\bs P_s}}}
			\sqrt{1\pm i{t-T\ov\sigma E\fn{\bs P_s}}\paren{1-\bs v^2\fn{\bs P_s}}}
			},
}
for any polynomial $Q\fn{\bs p}$. Note that the Gaussian integral can be performed when
\al{
1
	&>	\mp\Im{t-T\ov \sigma E\fn{\bs P_s}}
	=	\pm{t-T\ov\sigma\ab{E\fn{\bs P_s}}}{\Im E\fn{\bs P_s}\ov\ab{E\fn{\bs P_s}}},
		\label{Gaussian condition}\\
1
	&>	\mp\Im{\paren{t-T}\paren{1-\bs v^2\fn{\bs P_s}}\ov \sigma E\fn{\bs P_s}}
	=	\pm{t-T\ov\sigma\ab{E\fn{\bs P_s}}}
			{\paren{1-\Re\bs v^2\fn{\bs P_s}}\Im E\fn{\bs P_s}
			-\Re E\fn{\bs P_s}\Im\bs v^2\fn{\bs P_s}
				\ov\ab{E\fn{\bs P_s}}}.
		\label{Gaussian condition 2}
}

To summarize, the saddle-point method yields
\al{
\mc U_{s,\sigma;X,\bs P}\fn{x}
	&=	
		\paren{\sigma\ov\pi}^{3/4}
		\paren{1\ov2E_\Psi\fn{\bs P_s}\sigma^3}^{1/2}
		{U\fn{s,\bs P_s}
			e^{
			-iE_\Psi\fn{\bs P_s}\paren{t-T}
			+i\bs P_s\cdot\paren{\bs x-\bs X}
			-{\sigma\ov2}\paren{\bs P_s-\bs P}^2
			}	
		\ov
			\paren{1\pm i{t-T\ov \sigma E_\Psi\fn{\bs P_s}}}
			\sqrt{1\pm i{t-T\ov\sigma E_\Psi\fn{\bs P_s}}\paren{1-\bs v^2\fn{\bs P_s}}}}
		,\\
\mc V_{s,\sigma;X,\bs P}\fn{x}
	&=	
		\paren{\sigma\ov\pi}^{3/4}
		\paren{1\ov2E_\Psi\fn{\bs P_s}\sigma^3}^{1/2}
		{V\fn{s,\bs P_s}
			e^{
			iE_\Psi\fn{\bs P_s}\paren{t-T}
			-i\bs P_s\cdot\paren{\bs x-\bs X}
			-{\sigma\ov2}\paren{\bs P_s-\bs P}^2
			}\ov
			\paren{1\pm i{t-T\ov \sigma E_\Psi\fn{\bs P_s}}}
			\sqrt{1\pm i{t-T\ov\sigma E_\Psi\fn{\bs P_s}}\paren{1-\bs v^2\fn{\bs P_s}}}}
		.	\label{saddle point full}
}
When necessary we may expand them using
\al{
\paren{E\fn{\bs P+\Delta\bs P}}^{-1/2}
	&=	\paren{m^2+\paren{\bs P+\Delta\bs P}^2}^{-1/4}
	=	\paren{m^2+\bs P^2+2\bs P\cdot\Delta\bs P+\cdots}^{-1/4}\nn
	&=	\paren{E\fn{\bs P}}^{-1/2}\paren{1-{\bs v\fn{\bs P}\cdot\Delta\bs P\ov2E\fn{\bs P}}+\cdots},
}
etc., and the leading order result for the large $\sigma$ limit is\footnote{
We have taken up to the $\sigma^{-1}$ order in the exponent since the terms of order $\sigma^0$ are pure imaginary and just give a phase factor.
}
\al{
\mc U_{s,\sigma;X,\bs P}\fn{x}
	&=	
		\paren{\sigma\ov\pi}^{3/4}\paren{2\pi\ov\sigma}^{3/2}
		\paren{1\ov\paren{2\pi}^32E_\Psi\fn{\bs P}}^{1/2}
		U\fn{s,\bs P}
			e^{
			-iE_\Psi\fn{\bs P}\paren{t-T}
			+i\bs P\cdot\paren{\bs x-\bs X}
			-{\paren{\bs x-\bs \Xi\fn{t}}^2\ov2\sigma}
			}
		,\\
\mc V_{s,\sigma;X,\bs P}\fn{x}
	&=	
		\paren{\sigma\ov\pi}^{3/4}\paren{2\pi\ov\sigma}^{3/2}
		\paren{1\ov\paren{2\pi}^32E_\Psi\fn{\bs P}}^{1/2}
		V\fn{s,\bs P}
			e^{
			iE_\Psi\fn{\bs P}\paren{t-T}
			-i\bs P\cdot\paren{\bs x-\bs X}
			-{\paren{\bs x-\bs \Xi\fn{t}}^2\ov2\sigma}
			}
		.
}
This is used in Eqs.~\eqref{plane wave limit} and \eqref{vector saddle point}.

In the large $\sigma$ limit, we may iteratively solve the saddle point condition~\eqref{saddle point condition} by $\bs P_s=\bs P+\Delta_1\bs P+\Delta_2\bs P+\cdots$ with $\Delta_n\bs P=\Or{\sigma^{-n}}$. The result is
\al{
\bs P_s
	&=	\bs P\pm i{\bs x-\bs \Xi\fn{t}\ov\sigma}+\Or{1\ov\sigma^2},
		\label{Ps is P plus delta}
}
where $\bs \Xi\fn{t}:=\bs X+\bs V\paren{t-T}$ with $\bs V:=\bs v\fn{\bs P}$, corresponding to Eq.~\eqref{delta x defined}.
The zeroth and second derivatives read\footnote{
We may also rewrite
\als{
F_\pm\fn{\bs P_s}
	&=\mp i{m^2\ov E\fn{\bs P}}\paren{t-T}
		\pm i\bs P\cdot\paren{\bs x-\bs \Xi\fn{t}}
		-{\paren{\bs x-\bs \Xi\fn{t}}^2\ov2\sigma}
		+\Or{1\ov\sigma^2}\nn
	&=\mp i{m^2\ov E\fn{\bs P}}\paren{t-T}
		-{\sigma\ov2}\bs P^2
		-{\paren{\bs x-\bs \Xi\fn{t}\mp i\sigma\bs P}^2\ov2\sigma}
		+\Or{1\ov\sigma^2}.
}
}
\al{
F_\pm\fn{\bs P_s}
	&=	\mp iE\fn{\bs P}\paren{t-T}
		\pm i\bs P\cdot\paren{\bs x-\bs X}
		-{\paren{\bs x-\bs \Xi\fn{t}}^2\ov2\sigma}
		+\Or{1\ov\sigma^2}
		,\\
{\p^2F_\pm\fn{\bs P_s}\ov\p p_i\p p_j}
	&=	-\sigma\delta_{ij}\mp i{t-T\ov E\fn{\bs P}}\paren{\delta_{ij}-V_iV_j}
		+\Or{1\ov\sigma},
}
where we used
\al{
E\fn{\bs P_s}
	&=	E\fn{\bs P}\pm i{\bs V\cdot\paren{\bs x-\bs \Xi\fn{t}}\ov\sigma}+\Or{1\ov\sigma^2},\\
\bs v\fn{\bs P_s}
	&=	\bs V
		\pm i{\bs x-\bs \Xi\fn{t}-\bs V\sqbr{\bs V\cdot\paren{\bs x-\bs \Xi\fn{t}}}\ov\sigma E\fn{\bs P}}+\Or{1\ov\sigma^2}.
}
Especially, the necessary conditions~\eqref{Gaussian condition} and \eqref{Gaussian condition 2} read, at the leading order,
\al{
1	&>	{t-T\ov\sigma^2}{\bs V\cdot\paren{\bs x-\bs \Xi\fn{t}}\ov \bs P^2+m^2},\\
1	&>	-{t-T\ov\sigma^2}{
			\bs V\cdot\paren{\bs x-\bs \Xi\fn{t}}
		\ov \bs P^2+m^2}\paren{1-\bs V^2}.
}

\section{Wave and particle limits for decaying particle}\label{limits section}

\subsection{Wave limit}\label{wave limit section}
In the wave limit of the initial state, $\sigma_0\gg\sigma_a$, we get
\al{
\sigma_s
	&\to	{\sigma_1\sigma_2\ov\sigma_1+\sigma_2}
	=:		\sigma_\tx{out},
}
and then Eqs.~\eqref{Delta V squared}--\eqref{explicit R} reduce to
\al{
\Delta\bs V^2
	&\to	{\sigma_\tx{out}\ov\sigma_1+\sigma_2}
			\paren{\bs V_1-\bs V_2}^2,\\
\sigma_t
	&\to	{\sigma_1+\sigma_2\ov\paren{\bs V_1-\bs V_2}^2},\\
\mf T
	&\to	-{\paren{\bs V_1-\bs V_2}\cdot\paren{\bs{\mf X}_1-\bs{\mf X}_2}
				\ov
				\paren{\bs V_1-\bs V_2}^2},\\
\mc R
	&\to	{1\ov\sigma_1+\sigma_2}
			\sqbr{
				\paren{\bmf X_1-\bmf X_2}^2
				-{\sqbr{
					\paren{\bs V_1-\bs V_2}\cdot\paren{\bmf X_1-\bmf X_2}
					}^2\ov\paren{\bs V_1-\bs V_2}^2}
			},\\
\intertext{where we used, for arbitrary $\bs Q_A$ and $\bs Q_A'$,}
\ol{\bs Q\cdot\bs Q'}-\ol{\bs Q}\cdot\ol{\bs Q'}
	&\to	{\sigma_\tx{out}\ov\sigma_1+\sigma_2}
			\paren{\bs Q_1-\bs Q_2}\cdot\paren{\bs Q_1'-\bs Q_2'}.
}
(Recall that $\bmf X_1-\bmf X_2=\bs X_1-\bs X_2-\bs V_1T_1+\bs V_2T_2$.)
Note that the $\bs V_0$ dependence drops out in the wave limit.\footnote{
Note however that the $\bs V_0$ dependence in the zero eigenvector~\eqref{zero eigenvector given} still remains; see footnote~\ref{zero eigenvector}.
}

In the limit, the eigenvalues~\eqref{eigenvalues of M} become
\al{
&{2\ov\sigma_1+\sigma_2},&
&{1\ov2\sigma_0},&
&{1\ov\sigma_0}{\paren{\bs V_1-\bs V_0}^2+\paren{\bs V_2-\bs V_0}^2\ov
	\paren{\bs V_1-\bs V_2}^2}
		,
}
where the first two are two-fold degenerate per each.

\subsection{Particle limit}
In the particle limit of the initial state $\sigma_0\ll\sigma_a$, we obtain
\al{
\sigma_s
	&\to	\sigma_0,\\
\Delta\bs V^2
	&\to	\sigma_0\paren{
				{\paren{\delta\bs V_1}^2\ov\sigma_1}
				+{\paren{\delta\bs V_2}^2\ov\sigma_2}
				},\\
\sigma_t
	&\to	\paren{
				{\paren{\delta\bs V_1}^2\ov\sigma_1}
				+{\paren{\delta\bs V_2}^2\ov\sigma_2}
				}^{-1},\\
\mf T
	&\to	-{{\delta\bmf X_1\cdot\delta\bs V_1\ov\sigma_1}+{\delta\bmf X_2\cdot\delta\bs V_2\ov\sigma_2}\ov{\paren{\delta\bs V_1}^2\ov\sigma_1}
				+{\paren{\delta\bs V_2}^2\ov\sigma_2}},\\
\mc R
	&\to	{\paren{\delta\bmf X_1}^2\ov\sigma_1}
			+{\paren{\delta\bmf X_2}^2\ov\sigma_2}
			-{\paren{{\delta\bmf X_1\cdot\delta\bs V_1\ov\sigma_1}+{\delta\bmf X_2\cdot\delta\bs V_2\ov\sigma_2}}^2
			\ov{\paren{\delta\bs V_1}^2\ov\sigma_1}
				+{\paren{\delta\bs V_2}^2\ov\sigma_2}},\\
\intertext{where we used, for arbitrary $\bs Q_A$ and $\bs Q_A'$,}
\ol{\bs Q\cdot\bs Q'}-\ol{\bs Q}\cdot\ol{\bs Q'}
	&\to	\sigma_0\paren{
				{\delta\bs Q_1\cdot\delta\bs Q_1'\ov\sigma_1}
				+{\delta\bs Q_2\cdot\delta\bs Q_2'\ov\sigma_2}
				}.
}
(Recall that $\delta\bs V_a=\bs V_a-\bs V_0$ and that $\delta\bmf X_a=\bs X_a-\bs X_0-T_a\bs V_a+T_0\bs V_0$.)
The eigenvalues~\eqref{eigenvalues of M} become
\al{
&{1\ov\sigma_1},&
&{1\ov\sigma_2},&
&{1\ov\sigma_1\sigma_2}
	{\paren{\delta\bs V_1}^2+\paren{\delta\bs V_2}^2\ov
		{\paren{\delta\bs V_1}^2\ov\sigma_1}
		+{\paren{\delta\bs V_2}^2\ov\sigma_2}},
}
where the first two are two-fold degenerate per each.

More concretely,
\al{
\mf T
	&=		-{{\paren{\bs X_1-\bs X_0-\bs V_1T_1+\bs V_0T_0}\cdot\paren{\bs V_1-\bs V_0}\ov\sigma_1}
			+{\paren{\bs X_2-\bs X_0-\bs V_2T_2+\bs V_0T_0}\cdot\paren{\bs V_2-\bs V_0}\ov\sigma_2}\ov
				{\paren{\bs V_1-\bs V_0}^2\ov\sigma_1}
				+{\paren{\bs V_2-\bs V_0}^2\ov\sigma_2}},\\
\mc R
	&=	{\paren{\bs X_1-\bs X_0-\bs V_1T_1+\bs V_0T_0}^2\ov\sigma_1}
			+{\paren{\bs X_2-\bs X_0-\bs V_2T_2+\bs V_0T_0}^2\ov\sigma_2}\nn
	&\quad
			-{\paren{{\paren{\bs X_1-\bs X_0-\bs V_1T_1+\bs V_0T_0}\cdot\paren{\bs V_1-\bs V_0}\ov\sigma_1}
				+{\paren{\bs X_2-\bs X_0-\bs V_2T_2+\bs V_0T_0}\cdot\paren{\bs V_2-\bs V_0}\ov\sigma_2}}^2
			\ov{\paren{\bs V_1-\bs V_0}^2\ov\sigma_1}
				+{\paren{\bs V_2-\bs V_0}^2\ov\sigma_2}}.
}
Without loss of generality, we may set $X_0=\paren{T_0,\bs X_0}=0$, and then we obtain
\al{
\mf T
	&=		-{{\paren{\bs X_1-\bs V_1T_1}\cdot\paren{\bs V_1-\bs V_0}\ov\sigma_1}+{\paren{\bs X_2-\bs V_2T_2}\cdot\paren{\bs V_2-\bs V_0}\ov\sigma_2}\ov
				{\paren{\bs V_1-\bs V_0}^2\ov\sigma_1}
				+{\paren{\bs V_2-\bs V_0}^2\ov\sigma_2}},\\
\mc R
	&=	{\paren{\bs X_1-\bs V_1T_1}^2\ov\sigma_1}
			+{\paren{\bs X_2-\bs V_2T_2}^2\ov\sigma_2}
		-{\paren{{\paren{\bs X_1-\bs V_1T_1}\cdot\paren{\bs V_1-\bs V_0}\ov\sigma_1}
				+{\paren{\bs X_2-\bs V_2T_2}\cdot\paren{\bs V_2-\bs V_0}\ov\sigma_2}}^2
			\ov{\paren{\bs V_1-\bs V_0}^2\ov\sigma_1}
				+{\paren{\bs V_2-\bs V_0}^2\ov\sigma_2}}.
}

\subsection{Decay at rest}
Finally, we list the corresponding expression to Eqs.~\eqref{Delta V squared}--\eqref{explicit R} for the decay at rest $\bs V_0=0$ (and hence $\bs P_0=0$ and $E_0=m_0$), without taking any limit:
\al{
\Delta\bs V^2
	&=	\sigma_s^2
		\sqbr{
			{\bs V_1^2\ov\sigma_0\sigma_1}
			+{\bs V_2^2\ov\sigma_0\sigma_2}
			+{\paren{\bs V_1-\bs V_2}^2\ov\sigma_1\sigma_2}
			},\\
\sigma_t
	&=	{1\ov \sigma_s}\sqbr{
			{\bs V_1^2\ov\sigma_0\sigma_1}
			+{\bs V_2^2\ov\sigma_0\sigma_2}
			+{\paren{\bs V_1-\bs V_2}^2\ov\sigma_1\sigma_2}
				}^{-1}, \\
\mf T
	&=		-{{\paren{\bs X_1-\bs X_0-\bs V_1T_1}\cdot\bs V_1\ov\sigma_1}
			+{\paren{\bs X_2-\bs X_0-\bs V_2T_2}\cdot\bs V_2\ov\sigma_2}\ov
				{\bs V_1^2\ov\sigma_1}
				+{\bs V_2^2\ov\sigma_2}},
					\label{curly T explicit}\\
\mc R
	&=	{\paren{\bs X_1-\bs X_0-\bs V_1T_1}^2\ov\sigma_1}
			+{\paren{\bs X_2-\bs X_0-\bs V_2T_2}^2\ov\sigma_2}\nn
	&\quad
			-{\paren{{\paren{\bs X_1-\bs X_0-\bs V_1T_1}\cdot\bs V_1\ov\sigma_1}
				+{\paren{\bs X_2-\bs X_0-\bs V_2T_2}\cdot\bs V_2\ov\sigma_2}}^2
			\ov{\bs V_1^2\ov\sigma_1}
				+{\bs V_2^2\ov\sigma_2}}.
				\label{curly R explicit}
}
An experimentalist-friendly parametrization for the decay at rest might be
\al{
\paren{\delta\bs P}^2
	&=	p_1^2+2p_1p_2\cos\theta+p_2^2,
		\label{delta P at rest}\\
\delta\omega
	&=	E_1+E_2-m_0
		-{\sigma_sp_1\ov\sigma_1E_1}\paren{p_1+p_2\cos\theta}
		-{\sigma_sp_2\ov\sigma_2E_2}\paren{p_2+p_1\cos\theta},\\
\sigma_t
	&=	{\sigma_0\sigma_1\sigma_2\ov\sigma_s}
		{1\ov
			\paren{\sigma_0+\sigma_1}{p_2^2\ov E_2^2}
			-2\sigma_0{p_1p_2\ov E_1E_2}\cos\theta
			+\paren{\sigma_0+\sigma_2}{p_1^2\ov E_1^2}
		},\label{sigma t at rest}
}
where
$p_a:=\ab{\bs P_a}$ for $a=1,2$;
the angle is defined by $\cos\theta:={\bs P_1\cdot\bs P_2\ov p_1 p_2}$;
and $E_a$ and $\sigma_s$ are given in Eqs.~\eqref{on shell energy} and \eqref{sigma s defined}, respectively.
One may further take the above plane-wave or particle limit to simplify the expression if one wishes.

Without loss of generality, we may set $X_0=\paren{T_0,\bs X_0}=0$, and then Eqs.~\eqref{curly T explicit} and \eqref{curly R explicit} further simplify to
\al{
\mf T
	&=		-{{\paren{\bs X_1-\bs V_1T_1}\cdot\bs V_1\ov\sigma_1}
			+{\paren{\bs X_2-\bs V_2T_2}\cdot\bs V_2\ov\sigma_2}\ov
				{\bs V_1^2\ov\sigma_1}
				+{\bs V_2^2\ov\sigma_2}},\\
\mc R
	&=	{\paren{\bs X_1-\bs V_1T_1}^2\ov\sigma_1}
		+{\paren{\bs X_2-\bs V_2T_2}^2\ov\sigma_2}
		-{\paren{{\paren{\bs X_1-\bs V_1T_1}\cdot\bs V_1\ov\sigma_1}
				+{\paren{\bs X_2-\bs V_2T_2}\cdot\bs V_2\ov\sigma_2}}^2
			\ov{\bs V_1^2\ov\sigma_1}
				+{\bs V_2^2\ov\sigma_2}}.
}

To cultivate intuition, we present the results for a simple configuration $\sigma_2=\sigma_1$, $\bs V_0=0$, $\bs V_2=-\bs V_1$, and $T_1=T_2=:T_\tx{out}$:
\al{
\mf T
	&=	T_\tx{out}-{\bs X_1-\bs X_2\ov2\ab{\bs V_1}}\cdot{\bs V_1\ov\ab{\bs V_1}},\\
\mc R
	&=	{1\ov\sigma_1}\paren{
			\bs X_1^2+\bs X_2^2
			-{1\ov2}\sqbr{{\bs V_1\ov\ab{\bs V_1}}\cdot\paren{\bs X_1-\bs X_2}}^2
				}.
}

\section{Boundary contributions in boundary limit}\label{Boundary limit}
Here we present the boundary contributions in boundary limit $\ab{\mf T-T_\tx{in/out}}\ll\sqrt{\sigma_t}$, which might be applicable for $\sigma_t\paren{\delta\omega}^2\lesssim1$ too.

As discussed in the paragraph containing Eq.~\eqref{opposite limit}, the expression~\eqref{dP limit taken} is valid only when $\sigma_t\paren{\delta\omega}^2\gg1$ at $\mf T=T_\tx{in/out}$.
It might be convenient if we have an expression in the boundary limit $\ab{\mf T-T_\tx{in/out}}\ll\sqrt{\sigma_t}$, valid for small $\delta\omega$ too.
For that purpose, we expand the rational function in Eq.~\eqref{G limit} around $\mf T-T_\tx{in/out}=0$ and \emph{naively} replace, by using Eq.~\eqref{Dawson expanded}, as
\al{
{2\ov\pi}{1\ov\sigma_t\paren{\delta\omega}^2}
	&\mapsto	\sqbr{{4\ov\pi}F^2\fn{\sqrt{\sigma_t\ov2}\delta\omega}+e^{-\sigma_t\paren{\delta\omega}^2}},
}
to obtain
\al{
e^{-\sigma_t\paren{\delta\omega}^2}\ab{G\fn{\mf T}}^2
	&\simeq
		e^{-{\paren{\mf T-T_\tx{in/out}}^2\ov\sigma_t}}\sqbr{{4\ov\pi}F^2\fn{\sqrt{\sigma_t\ov2}\delta\omega}+e^{-\sigma_t\paren{\delta\omega}^2}}.
}
The first and second terms in the square brackets are exactly the dashed and dot-dashed lines, respectively, in Fig.~\ref{imaginary fig} (and their sum is the solid line)\red{.}
For reference, we show the boundary contribution with this naive replacement:
\al{
\df P
	&\simeq	
		{\kappa^2\ov2}
		{1\ov2E_0}
		{\df^3\bs P_1\ov\paren{2\pi}^{3}2E_1}
		{\df^3\bs P_2\ov\paren{2\pi}^{3}2E_2}
		\paren{2\pi}^4
		\paren{\sqrt{\sigma_t\ov\pi}}
		\paren{\paren{\sigma_s\ov\pi}^{3/2}e^{-\sigma_s\paren{\delta\bs P}^2}}
			\nn
	&\qquad\times
		\sqrt{
		{\sigma_t\ov\pi^5}
		\paren{\sigma_s\ov\sigma_0\sigma_1\sigma_2}^3}\,
		\df^3\bs X_1\,\df^3\bs X_2\,
		e^{-\mc R}
		e^{-{\paren{\mf T-T_\tx{in/out}}^2\ov\sigma_t}}
		\sqbr{{4\ov\pi}F^2\fn{\sqrt{\sigma_t\ov2}\delta\omega}+e^{-\sigma_t\paren{\delta\omega}^2}}.
}
The formal limit~\eqref{delta function limit} of this expression reads 
\al{
\df P
	&\to
		{\kappa^2\ov2}
		{1\ov2E_0}
		{\df^3\bs P_1\ov\paren{2\pi}^{3}2E_1}
		{\df^3\bs P_2\ov\paren{2\pi}^{3}2E_2}
		\paren{2\pi}^4
		\paren{\sqrt{\sigma_t\ov\pi}}
		\paren{\paren{\sigma_s\ov\pi}^{3/2}e^{-\sigma_s\paren{\delta\bs P}^2}}
			\nn
	&\qquad\times
		\sqrt{
		{\sigma_t\ov\pi^5}
		\paren{\sigma_s\ov\sigma_0\sigma_1\sigma_2}^3}\,
		\df^6y\,
		e^{-\mc R}\,
		\sqrt{\pi\sigma_t}\,\delta\fn{\mf T-T_\tx{in/out}}
			\sqbr{{4\ov\pi}F^2\fn{\sqrt{\sigma_t\ov2}\delta\omega}+e^{-\sigma_t\paren{\delta\omega}^2}}.
}
\emph{Naively,} integration over $\df^5y\,e^{-\mc R}$ would again give Eq.~\eqref{five dimensional integral}, and then the $y_0$ integral over the delta function gives the extra factor $\sqrt{\paren{\delta\bs V_1}^2+\paren{\delta\bs V_2}^2}$:
\al{
\df P
	&=	{\kappa^2\ov2}
		{1\ov2E_0}
		{\df^3\bs P_1\ov\paren{2\pi}^{3}2E_1}
		{\df^3\bs P_2\ov\paren{2\pi}^{3}2E_2}
		\paren{2\pi}^4
		\paren{\paren{\sigma_s\ov\pi}^{3/2}e^{-\sigma_s\paren{\delta\bs P}^2}}
			\sigma_t\sqbr{{4\ov\pi}F^2\fn{\sqrt{\sigma_t\ov2}\delta\omega}+e^{-\sigma_t\paren{\delta\omega}^2}}.
}
Recall that $\delta\bs P$, $\delta\omega$, and $\sigma_t$ simplify to Eqs.~\eqref{delta P at rest}--\eqref{sigma t at rest} for the case of the decay at rest.

\bibliographystyle{utphys}
\bibliography{refs}

\providecommand{\href}[2]{#2}\begingroup\raggedright\begin{thebibliography}{10}

\bibitem{Peskin:1995ev}
M.~E. Peskin and D.~V. Schroeder, {\em {An Introduction to quantum field
  theory}}.
\newblock Addison-Wesley, Reading, USA, 1995.
\newblock
\url{http://www.slac.stanford.edu/~mpeskin/QFT.html}.
\newblock

\bibitem{Weinberg:1995mt}
S.~Weinberg, {\em {The Quantum theory of fields. Vol. 1: Foundations}}.
\newblock Cambridge University Press,
2005.
\newblock

\bibitem{Ishikawa:2005zc}
K.~Ishikawa and T.~Shimomura, ``{Generalized S-matrix in mixed
  representations},'' \href{http://dx.doi.org/10.1143/PTP.114.1201}{{\em Prog.
  Theor. Phys.} {\bfseries 114} (2006) 1201--1234},
\href{http://arxiv.org/abs/hep-ph/0508303}{{\ttfamily arXiv:hep-ph/0508303
  [hep-ph]}}.

\bibitem{doi:10.1143/PTP.17.419}
H.~Araki, Y.~Munakata, M.~Kawaguchi, and T.~Got{\^o}, ``Quantum field theory of
  unstable particles,'' \href{http://dx.doi.org/10.1143/PTP.17.419}{{\em
  Progress of Theoretical Physics} {\bfseries 17} no.~3, (1957) 419--442}.
  \url{http://dx.doi.org/10.1143/PTP.17.419}.

\bibitem{doi:10.1143/PTP.18.101b}
H.~Araki, Y.~Munakata, M.~Kawaguchi, and T.~Got{\^o}, ``Quantum field theory of
  unstable particles,'' \href{http://dx.doi.org/10.1143/PTP.18.101b}{{\em
  Progress of Theoretical Physics} {\bfseries 18} no.~1, (1957) 101}.
  \url{http://dx.doi.org/10.1143/PTP.18.101b}.

\bibitem{doi:10.1143/PTP.20.857}
K.~Yamamoto, ``Theory of unstable particles in the wave-packet-formalism,''
  \href{http://dx.doi.org/10.1143/PTP.20.857}{{\em Progress of Theoretical
  Physics} {\bfseries 20} no.~6, (1958) 857--867}.
  \url{http://dx.doi.org/10.1143/PTP.20.857}.

\bibitem{Cho:1993xe}
G.-C. Cho, H.~Kasari, and Y.~Yamaguchi, ``{The Time evolution of unstable
  particles},''
\href{http://dx.doi.org/10.1143/PTP.90.803}{{\em Prog. Theor. Phys.} {\bfseries
  90} (1993) 803--816}.

\bibitem{Ishikawa:2014eoa}
K.~Ishikawa and Y.~Tobita, ``{Matter-enhanced transition probabilities in
  quantum field theory},''
  \href{http://dx.doi.org/10.1016/j.aop.2014.02.007}{{\em Annals Phys.}
  {\bfseries 344} (2014) 118--178},
\href{http://arxiv.org/abs/1206.2593}{{\ttfamily arXiv:1206.2593 [hep-ph]}}.

\bibitem{Ishikawa:2013kba}
K.~Ishikawa and Y.~Tobita, ``{Finite-size corrections to Fermi's golden rule:
  I. Decay rates},'' \href{http://dx.doi.org/10.1093/ptep/ptt049}{{\em PTEP}
  {\bfseries 2013} (2013) 073B02},
\href{http://arxiv.org/abs/1303.4568}{{\ttfamily arXiv:1303.4568 [hep-ph]}}.

\bibitem{Ishikawa:2016lnn}
K.~Ishikawa and Y.~Tobita, ``{Finite-size corrections to Fermi's Golden rule
  II: Quasi-stationary composite states},''
\href{http://arxiv.org/abs/1607.08522}{{\ttfamily arXiv:1607.08522 [hep-ph]}}.

\bibitem{Ishikawa:2014uma}
K.~Ishikawa, T.~Tajima, and Y.~Tobita, ``{Anomalous radiative transitions},''
  \href{http://dx.doi.org/10.1093/ptep/ptu168}{{\em PTEP} {\bfseries 2015}
  (2015) 013B02},
\href{http://arxiv.org/abs/1409.4339}{{\ttfamily arXiv:1409.4339 [hep-ph]}}.

\bibitem{Stueckelberg:1951zz}
E.~C.~G. Stueckelberg, ``{Relativistic Quantum Theory for Finite Time
  Intervals},''
\href{http://dx.doi.org/10.1103/PhysRev.81.130}{{\em Phys. Rev.} {\bfseries 81}
  (1951) 130--133}.

\bibitem{Weisskopf:1930au}
V.~Weisskopf and E.~P. Wigner, ``{Calculation of the natural brightness of
  spectral lines on the basis of Dirac's theory},''
\href{http://dx.doi.org/10.1007/BF01336768}{{\em Z. Phys.} {\bfseries 63}
  (1930) 54--73}.

\bibitem{Ishikawa:2014ela}
K.~Ishikawa, T.~Nozaki, M.~Sentoku, and Y.~Tobita, ``{Transition Probability
  for the Neutrino Wave in Muon Decay and Oscillation Experiments},''
\href{http://arxiv.org/abs/1405.0582}{{\ttfamily arXiv:1405.0582 [hep-ph]}}.

\bibitem{Newton:1949cq}
T.~D. Newton and E.~P. Wigner, ``{Localized States for Elementary Systems},''
\href{http://dx.doi.org/10.1103/RevModPhys.21.400}{{\em Rev. Mod. Phys.}
  {\bfseries 21} (1949) 400--406}.

\end{thebibliography}\endgroup

\end{document}